\documentclass[journal]{IEEEtran}
\usepackage{cite}
\usepackage{amsmath,amssymb,amsfonts}
\usepackage{graphicx}
\usepackage{subcaption}
\usepackage{wrapfig}
\usepackage{multirow}
\usepackage{placeins}
\usepackage{booktabs}
\usepackage{algorithm2e}
\usepackage{float}
\usepackage{stfloats} 
\usepackage{xcolor}
\usepackage{tikz}
\usetikzlibrary{arrows.meta,positioning,calc,fit}

\usepackage[colorlinks=true, linkcolor=blue, citecolor=blue, urlcolor=blue]{hyperref}

\renewcommand{\footnoterule}{
  \kern -3pt 
  \hrule width 0.5\textwidth height 0.4pt 
  \kern 2.6pt 
}

\title{Observer-usable Information as a Task-specific Image Quality Metric}

\author{Changjie~Lu*,~\IEEEmembership{}
        Sourya Sengupta*,
        Hua~Li,~\IEEEmembership{Senior Member},
        ~Mark~A.~Anastasio, ~\IEEEmembership{Fellow}
\thanks{*C. Lu and S. Sengupta contributed equally to this work.}
\thanks{C. Lu and M. Anastasio are with the Department of Bioengineering, University of Illinois Urbana-Champaign, Urbana, IL, USA.}
\thanks{S. Sengupta is with the Department of Electrical Engineering. \& Computer Eng., University of Illinois Urbana-Champaign, Urbana, IL, USA}
\thanks{H. Li is with the Department of Radiation Oncology at Washington University in St. Louis, MO, USA. She is also affiliated with the Department of Bioengineering, University of Illinois Urbana-Champaign, Urbana, IL, USA.}
\thanks{Correspondence should be addressed to Mark A. Anastasio (email: maa@illinois.edu) and Hua Li (email: li.hua@wustl.edu).}\\
[-6ex]
}

\begin{document} 
\maketitle


\begin{abstract}
Objective, task-based measures of image quality (IQ) have been widely advocated for assessing and optimizing medical imaging technologies. Besides signal detection theory-based measures, information-theoretic quantities have been proposed to quantify task-based IQ. For example, task-specific information (TSI), defined as the mutual information between an image and a task variable, represents an optimal measure of how informative an image is for performing a specified task. However, like the ideal observer from signal detection theory,  TSI  does not quantify the amount of
task-relevant information in an image that can be exploited by a sub-ideal observer.
 A recently proposed relaxation of TSI, termed predictive $\mathcal{V}$-information ($\mathcal{V}$-info), removes this limitation and can quantify the utility of an image with consideration of a specified family of sub-ideal observers.  {In this study, for the first time, we introduce and investigate $\mathcal{V}$-info  as an objective, task-specific IQ metric.} To corroborate its usefulness, a stylized magnetic resonance image restoration problem is considered in which $\mathcal{V}$-info is employed to quantify signal detection or discrimination performance. 
The presented experiments show that, for binary classification tasks, 
$\mathcal{V}$-info varies consistently with the area under the receiver 
operating characteristic (ROC) curve in regimes where class separability 
changes with observer capacity or imaging conditions. However, unlike AUC, 
$\mathcal{V}$-info remains sensitive in regimes where discrimination 
performance approaches saturation. In addition, $\mathcal{V}$-info is readily 
applicable to multi-class ($>2$) tasks where ROC analysis is less natural. 
These findings suggest that $\mathcal{V}$-info can serve as a complementary 
task-based image quality measure alongside traditional signal detection 
theory–based metrics.
\end{abstract}
\begin{IEEEkeywords}
 Objective image quality assessment, numerical observers, information theory
\end{IEEEkeywords}

\vspace{-0.1in}
\section{INTRODUCTION}

The evaluation of medical image quality (IQ) plays a central role in the development and refinement of imaging systems and image processing algorithms~\cite{swets1979roc, maansson2000methods, wilson2013methodology,doi2006diagnostic,beutel2000handbook,cunningham1999signal,tapiovaara1993snr,vennart1997icru}. Conventional IQ metrics, such as peak signal-to-noise ratio (PSNR) and structural similarity index (SSIM)~\cite{ssim}, are frequently used to quantify fidelity between restored and original images. However, these metrics do not consistently reflect the utility of images for clinical decision-making~\cite{barrett2013foundations}. 
To address this, objective, task-based measures of image quality (IQ) have been widely advocated for assessing and optimizing medical imaging technologies\cite{barrett2013foundations,barrett1998objective,kupinski2003optimizing}.

The performance of numerical observers (NOs) on signal detection or discrimination tasks has been widely employed as objective task-based IQ measure in early-stage studies of imaging technologies~\cite{ barrett2015task,park2013model,krupinski2021important}.
Depending on the goal of the study, either ideal observers (IOs) or sub-ideal NOs can be employed.
The IO is a special NO that implements an optimal decision strategy, yielding the best possible performance on a detection or discrimination task.  One application of the IO involves the optimization of imaging hardware and data acquisition designs~\cite{zhou2008aperture, ghaly2015optimization,sidky2008depth, kupinski2003optimizing, li2025estimating}.
However, the IO can be difficult, or intractable, to compute\cite{kupinski2003ideal, abbey2008ideal, zhou2020approximating,he2008toward} in practice.  

 {To address this, sub-ideal NOs, which represent observers that cannot fully
exploit all task-relevant information in the image data, have been employed as
surrogates of the
IO\cite{sahu2006assessment,sanchez2014task,oldan2004channelized,sanchez2016use}.}
Sub-ideal NOs enable routine task-specific assessments and optimizations of imaging technologies under feasible computational constraints. Sub-ideal NOs yield measures of task performance for which only a restricted amount of statistical information in the image data is utilized.
Anthropomorphic NOs,
such as certain channelized Hotelling observers~\cite{yao1992predicting, wunderlich2009estimation, noferini2016ct, favazza2017use}, are a specific type of sub-ideal NO that aim to approximate human performance on the task. They can be employed, for example, to optimize image reconstruction methods that produce images intended for human reading.
 {Deep neural network-based observers~\cite{zhou2019approximating,zhou2020approximating}  represent another category  of sub-ideal NOs, as they generally possess limited model capacity and implement sub-optimal decision strategies.}

As an alternative to signal detection theory-based IQ measures, information-theoretic quantities have been proposed as task-based IQ measures\cite{metz1973evaluation, roc2, li2013relating, clarkson2020bayesian, clarkson2019relation, clarkson2010fisher, o2002information}. 
For example, Neifeld et al.\cite{neifeld2007task} defined task-specific information (TSI)  as the mutual information (MI) between  
imaging data and a task variable, which
represents a measure of the optimal utility of the data with consideration of the specified task~\cite{ashok2007task,neifeld2007task}. 
Clarkson and Cushing demonstrated that TSI and IO receiver operating characteristic (ROC) analyses are equivalent descriptions of the best possible performance of any observer on a binary signal detection task~\cite{si1}, and extended this equivalence to general multi-class discrimination tasks\cite{si2}. 
Like the IO performance, TSI cannot be increased via image processing due to the data processing inequality (DPI)~\cite{dpi}.
Accordingly, TSI can be considered as a task-based IQ measure whose domain of applicability is like that of the IO.
Similar to the IO, TSI is generally difficult, or intractable, to compute for large-scale problems.

A sub-ideal observer equivalent of TSI has not been available, which has limited its application in practice.
Excitingly, a recently proposed information-theoretic quantity called \emph{predictive $\mathcal{V}$-information} ($\mathcal{V}$-info)~\cite{xu2020theory} represents a relaxation of MI and addresses this need.
$\mathcal{V}$-information is defined with respect to a restricted family of functions $\mathcal{V}$
that map a random vector (e.g., an image) to a probability distribution over a target variable (e.g., vector of class labels). For example, this class of functions could correspond to those that
 can be implemented by a given probabilistic classifier. 
In this way, $\mathcal{V}$-info can explicitly incorporate capacity constraints associated with a class of sub-ideal observers. This allows it to  quantify the maximum amount  
of task-relevant information in a random vector that
is \emph{usable} by a specified sub-ideal observer.
Hereafter, task-related information refers to a measure of how informative an image is for predicting or inferring a task variable.\footnote{Specifically, task-related information refers to a measure of the statistical dependencies between the source and target variables that can be exploited by the observer (decision maker)  for the specified task.}
Unlike MI or its special case of TSI, $\mathcal{V}$-info can be increased by image processing and can be computed readily.
In the special case where $\mathcal{V}$ contains all possible mappings, $\mathcal{V}$-info reduces to classic TSI.

In this study, for the first time, $\mathcal{V}$-info is introduced and investigated
as an objective, task-specific IQ metric.
While the primary focus is on image-quality assessment, the proposed framework can be applied directly to raw measurement data to quantify the task-relevant information acquired by a computed imaging system that is accessible to an observer within a specified class, independent of reconstruction or post-processing.
To demonstrate the utility of $\mathcal{V}$-info, virtual imaging experiments are conducted that involve an image restoration problem.
The performance of  NOs using original, restored, and ground truth images for signal detection and discrimination tasks is quantified by use of conventional task-based IQ measures and $\mathcal{V}$-info. The relative behavior of these alternative measures of IQ is then empirically studied.
The immediate applicability of $\mathcal{V}$-info to multi-class ($>$2) detection or discrimination problems, where ROC analysis is difficult to deploy, is also demonstrated.

The remainder of the paper is organized as follows. Section~\ref{background} contains a description of the general problem and salient mathematical definitions from information theory. $\mathcal{V}$-info is formally defined in Section~\ref{ouinfoasIQ} within the context of objective IQ assessment. Sections~\ref{application} and \ref{results} describe case studies, which demonstrate the effectiveness of $\mathcal{V}$-info as a task-based IQ metric when sub-ideal observers are relevant. Finally, a summary of the findings is presented in  Section~\ref{discussion}  along with a discussion of future research topics.

\section{Background}
\label{background}

\subsection{Formulation of Signal Detection Tasks}
\label{signaldetection}
A linear digital imaging system can be described as a continuous-to-discrete (C-D) mapping process:
\vspace{-0.03in}
\begin{equation}
    \textbf{g} = \mathcal{H}f(\textbf{r}) + \textbf{n},
\end{equation}

\noindent where \( \textbf{g} \in \mathbb{R}^{N \times 1} \) represents the measured image vector and \( f \) denotes the object function dependent on spatial coordinates \( \textbf{r} \in \mathbb{R}^{k \times 1} \), with \( k \geq 2 \). The operator $\mathcal{H}$ is a linear imaging operator that maps \( \mathbb{L}_2(\mathbb{R}^k) \) to \( \mathbb{R}^{N \times 1} \), and \( \textbf{n} \in \mathbb{R}^{N \times 1} \) represents the measurement noise. For simplicity, the object function \( f(\textbf{r}) \) will be denoted as \( \textbf{f} \) when its spatial dependence is not relevant. In practice, the elements of \(\textbf{g}\) are quantized into a finite set of discrete intensity levels. As such, in this work, \(\textbf{g}\) is considered to be a discrete random vector.

In an $\it{L}$-class signal detection task, the signal-absent and signal-present hypotheses can be described as\cite{green1966signal}:
\vspace{-0.03in}
\begin{subequations}
\begin{equation}
    \textbf{H}_{0} : \textbf{g} = \mathcal{H}(\textbf{f}_{\textbf{b}}) + \textbf{n} = \mathbf{b+n},
    \label{hypothesis-0}
\end{equation}
\begin{equation}
     \textbf{H}_{l} : \textbf{g} = \mathcal{H}(\textbf{f}_{\textbf{b}} + \textbf{f}_{\textbf{s}_{l}}) + \textbf{n} = \mathbf{b+s_{\it{l}}+n}.
    \label{hypothesis}
\end{equation}
\end{subequations}

\noindent Here, \(l = (1, ..., L-1)\) is an index and $\it{L}\geq 2$,
\( \textbf{f}_\textbf{b}\) and \( \textbf{f}_{\textbf{s}_{l}} \) denote the background and the $l$-th signal objects, respectively,
and \( \textbf{b} = \mathcal{H}\textbf{f}_{\textbf{b}} \) and \( \textbf{s} = \mathcal{H}\textbf{f}_{\textbf{s}_{l}} \) denote the corresponding images of the background and signal objects. 
The special case $L$=2 corresponds to a binary detection task.

\vspace{-0.1in}
\subsection{Mutual Information and Task-specific Information}
Mutual Information (MI) is a fundamental concept in information theory that measures how much knowing one random variable reduces uncertainty, i.e., Shannon entropy, about another\cite{shannon1948mathematical}.
Mathematically, the MI between two discrete random vectors \(\mathbf{X}\) and \(\mathbf{Y}\) is defined as\cite{shannon1948mathematical}:
\vspace{-0.05in}
\begin{equation}
I(\mathbf{Y}; \mathbf{X}) = H(\mathbf{Y}) - H(\mathbf{Y}\mid{\mathbf{X}}),
 \label{eq:MI-initial}
\end{equation}

\vspace{-0.05in}
\noindent where $H(\mathbf{Y})$ denotes the entropy of \(\mathbf{Y}\), and $H(\mathbf{Y}|\mathbf{X})$ denotes the conditional entropy of \(\mathbf{Y}\) given \(\mathbf{X}\). 
A larger MI value indicates that observations of $\mathbf{X}$ enable more accurate inference of $\mathbf{Y}$.
Formally, \eqref{eq:MI-initial} can be written as\cite{xu2020theory}:
\begin{equation}
\begin{aligned}
I(\mathbf{Y}; \mathbf{X}) &= H(\mathbf{Y}) - H(\mathbf{Y} \mid \mathbf{X}) \\
        &= H(\mathbf{Y}) - \mathbb{E}_{x,y \sim P_{\mathbf{X},\mathbf{Y}}}[-\log P_{\mathbf{Y} \mid \mathbf{X}}(y \mid x)],
\end{aligned}
\label{eq:MIdef}
\end{equation}

\noindent where \( P_{\mathbf{Y}|\mathbf{X}} \) represents the conditional probability distribution of \( \mathbf{Y} \) given \( \mathbf{X} \) and  \( P_{\mathbf{X},\mathbf{Y}} \) denotes the joint probability distribution.
The expectation operator $\mathbb{E}_{x,y \sim P_{\mathbf{X},\mathbf{Y}}}$ denotes the average taken over samples $(x, y)$ drawn from this joint distribution. 

Hereafter, a detection or discrimination task will be considered where \(\mathbf{X}\) and \(\mathbf{Y}\) denote the random image data and vector of categorical class labels. In this case, MI corresponds to the
previously proposed task-specific information (TSI) metric \cite{neifeld2007task}.
Let \(v[x](y)\) denote a mapping from a sample $x$ of \(\mathbf{X}\) to a probability distribution over \(\mathbf{Y}\). Such mappings are implemented in probabilistic classifiers or posterior probability models, such as neural networks that employ a softmax function at the output layer.
In this case,  \eqref{eq:MIdef} can  be expressed as \cite{cover1999elements}:
\begin{equation}
    I(\mathbf{Y}; \mathbf{X}) = H(\mathbf{Y}) - \inf_{v \in \mathcal{U}} \mathbb{E}_{x,y \sim P_{\mathbf{X},\mathbf{Y}}}[-\log v[x](y)],
\label{eq:MI}
\end{equation}

\noindent
where  \(\mathcal{U}\) denotes the set of all mappings from \(\mathbf{X}\) to probability distributions over \(\mathbf{Y}\).
By definition, \(\mathcal{U}\) also contains the mapping that is employed by the IO, $v[x](y)=P_{\mathbf{Y} \mid \mathbf{X}}(y \mid x)$, which achieves the infimum in \eqref{eq:MI}\cite{cover1999elements}.
Because of this, TSI is not influenced by the performance limitations of sub-ideal observers and is related to  IO performance.

Like the IO, TSI cannot be increased via image processing.  This is a consequence of the data processing inequality (DPI)~\cite{dpi}.  The DPI indicates that $ I(\mathbf{Y}; \textbf{M}) \leq I(\mathbf{Y}; \mathbf{X})$, where $\textbf{M}$ denotes any processed version of the image data $\mathbf{X}$.

\section{$\mathcal{V}$-information as an objective IQ metric}
\label{ouinfoasIQ}

\subsection{Definition of $\mathcal{V}$-information}

In practice, the approximation capacity\cite{baldi2019capacity} of network-based classifiers is limited.
Let  \(\mathcal{V}\subset \mathcal{U}\) denote the family of   mappings from an image to a  probability distribution over class labels that are achievable by such a non-ideal classifier.
As indicated by \eqref{eq:MI}, the definition of TSI is predicated upon the universal set of mappings $\mathcal{U}$; therefore, TSI only quantifies the amount of task-relevant information that can be used by an ideal classifier that possesses no capacity constraints.
In practice, there is an important need  to quantify the amount of task-relevant information that can actually be utilized by a non-ideal decision maker.

To address this, a relaxation of TSI termed \emph{predictive $\mathcal{V}$-information} ($\mathcal{V}$-info)\cite{xu2020theory} has been proposed.
Consider that $\mathcal{U}$ is replaced by $\mathcal{V}$ in the second term on the right-hand side of \eqref{eq:MI} and define:
\begin{equation}
   H_\mathcal{V}[\mathbf{Y}|\mathbf{X}] \equiv  \inf_{v \in \mathcal{V}} \mathbb{E}_{x,y \sim P_{\mathbf{X},\mathbf{Y}}}[-\log v[x](y)],
   \label{eq:conditional-v-entropy}
\end{equation}
which is known as the conditional $\mathcal{V}$-entropy\cite{xu2020theory}.
 {Here, the relaxation arises from replacing the unrestricted family of mappings
$\mathcal{U}$, over which the ideal observer is defined as the optimal decision
rule, with a restricted family of mappings $\mathcal{V}$ that represents a
specified class of non-ideal observers or predictors. Under this restriction,
$H_{\mathcal{V}}[\mathbf{Y} \vert \mathbf{X}]$ can be interpreted as a conditional
entropy that quantifies the residual uncertainty about the task variable
$\mathbf{Y}$ after observing the data $\mathbf{X}$ when prediction is constrained
to mappings in $\mathcal{V}$. In contrast, the true conditional entropy
$H(\mathbf{Y} \vert \mathbf{X})$ assumes an unrestricted observer and therefore
represents the residual uncertainty achievable by the ideal observer.}

This interpretation provides a natural bridge to a relaxed notion of
TSI. By quantifying how much the uncertainty
about $Y$ can be reduced by observing $X$ under the constraint that predictions
are formed using mappings in $\mathcal{V}$, one arrives naturally at the
definition of $\mathcal{V}$-info.
Specifically, the $\mathcal{V}$-info between $\mathbf{X}$ and $\mathbf{Y}$, denoted as $I_\mathcal{V}[\mathbf{X} \rightarrow \mathbf{Y}]$, is defined as\cite{xu2020theory,hewitt2021conditional}:
\footnote{In \cite{xu2020theory}, a null variable was introduced in the original definition of $\mathcal{V}$-info for technical reasons. The appendix of \cite{hewitt2021conditional} shows that this can be avoided.}
\begin{equation}
\begin{aligned}
I_\mathcal{V}[\mathbf{X} \rightarrow \mathbf{Y}] &\equiv H(\mathbf{Y}) - H_\mathcal{V}[\mathbf{Y}|\mathbf{X}].
\end{aligned}
\label{ouinfo}
\end{equation}

By definition, $\mathcal{V}$-info is a measure of
 the task-relevant predictive value
of the image $\mathbf{X}$ with consideration of a finite capacity classifier
 that can only implement the mappings in $\mathcal{V}$.
In this sense, $\mathcal{V}$-info can be interpreted as a measure of the maximum amount of task-relevant information in $\mathbf{X}$ that is \emph{usable} by  a classifier in $\mathcal{V}$ for predicting $\mathbf{Y}$.
Depending on the base of the logarithm employed, \(\mathcal{V}\)-info is measured in units of bits or nats.
Like MI, when $\mathbf{X}$ and $\mathbf{Y}$ are statistically independent, $I_\mathcal{V}[\mathbf{X} \rightarrow \mathbf{Y}]=0$, and $I_\mathcal{V}[\mathbf{X} \rightarrow \mathbf{Y}]$ takes on its maximum value when $\mathbf{X}$ fully determines $\mathbf{Y}$.  Additional mathematical properties
of $\mathcal{V}$-info can be found elsewhere\cite{xu2020theory}.
In the special case where \(\mathcal{V}\) = \(\mathcal{U}\), $\mathcal{V}$-info reduces to TSI.
To date, $\mathcal{V}$-info has been employed in machine learning-related studies that include dataset difficulty estimation via pointwise $\mathcal{V}$-info~\cite{ethayarajh2022understanding}, understanding useful representations~\cite{bakermakes,dubois2020learning}, and measuring large language model  performance~\cite{lu2023measuring}. Table \ref{tab:mi_vinfo_auc_comparison} summarizes the relationship between MI, $\mathcal{V}$-info, and other signal detection theory (SDT) metrics, such as AUC and accuracy.

\begin{table*}[t]
\centering
\caption{Comparison of Mutual Information, Predictive $\mathcal{V}$-Info, and signal detection theory (SDT) Metrics.}
\label{tab:mi_vinfo_auc_comparison}
\renewcommand{\arraystretch}{1.3}
\begin{tabular}{p{3.3cm} p{4.2cm} p{4.7cm} p{4.4cm}}
\toprule
\textbf{Property} & 
\textbf{Mutual Information $I(Y;X)$} & 
\textbf{Predictive $\mathcal{V}$-Info $I_{\mathcal{V}}(Y \rightarrow X)$} & 
\textbf{Canonical SDT Metrics} \\
\midrule

Underlying assumption &
Unrestricted ideal observer &
Observer optimal within a restricted class $\mathcal{V}$ &
Scalar decision variable with task-dependent variability\\

Interpretation &Total task-relevant statistical dependence between $X$ and $Y$; reduction in uncertainty of $Y$ given full access to joint distribution between $X$ and $Y$ &Observer-accessible task-relevant information; reduction in task uncertainty achievable by a predictor family $\mathcal{V}$ &Decision-level separability based on hard predictions; ignores posterior uncertainty and calibration \\


Decision rule dependence &
Independent of decision rules &
Independent of explicit decision rules (but is observer- and loss-aware) &
May depend on the chosen decision rule or operating point\\

Numerical value dependence &
Depends on the full joint distribution $p(X,Y)$ &
Depends on the optimal probabilistic predictor within $\mathcal{V}$ &
Depends only on the relative ranking of observer scores\\

Image processing effects &
Cannot be increased by invertible processing &
May increase under observer-matched (but not arbitrary) non-invertible processing &
May improve or degrade depending on how processing affects class separability \\

Limiting behavior &
Ideal-observer benchmark &
Converges to $I(Y;X)$ as $\mathcal{V}$ expands &
Saturates at perfect class separability \\

\bottomrule
\end{tabular}
\end{table*}

\vspace{-0.1in}
\subsection{$\mathcal{V}$-information for task-based IQ assessment}

$\mathcal{V}$-info holds significant potential as an information-theoretic measure of task-based IQ that quantifies the utility of an image with consideration of a class of sub-ideal observers. As overviewed above,  its properties are well-aligned for this purpose.
First, $\mathcal{V}$-info is  capable of being increased by image processing. This is consistent with the intuitive notion that certain image processing operations can   increase the utility of an image for sub-ideal observers, such as humans. For example, this is why image denoising methods are commonly applied in practice.
Second, $\mathcal{V}$-info is observer-specific and quantifies only task-relevant information that can be exploited by the specified class of observers.  This is consistent with the requirement that the utility of an image is dependent on the observer performing the task.
Finally, unlike TSI or IO performance, $\mathcal{V}$-info is computationally tractable.

Consider that a parameterized probabilistic sub-ideal NO (e.g.,  neural network-based) 
is to be employed to assess the quality of image data $\mathbf{X}$ based on a signal detection task. The family of mappings from samples of $\mathbf{X}$ to probability distributions over the class labels $\mathbf{Y}$
that are achievable by this NO will define the set $\mathcal{V}$.
The $\mathcal{V}$-info $I_\mathcal{V}[\mathbf{X} \rightarrow \mathbf{Y}]$ is a measure of the maximum amount of task-related information  present in $\mathbf{X}$ that can be exploited by the NO  for predicting  $\mathbf{Y}$. 
In this sense, $\mathcal{V}$-info can be viewed as an information-theoretic analog of signal detection performance measures of sub-ideal NO performance.

Because the relaxation in \eqref{eq:conditional-v-entropy} is defined for an arbitrary target random vector $\mathbf{Y}$, it remains applicable whether $\mathbf{Y}$ encodes a single multi-class label (e.g., one-hot vector) or multiple binary classification tasks (e.g., a vector of independent binary labels) \cite{song2020multi}. Specifically, in a classification task with $L$ possible classes, where $\mathbf{Y}$ indicates the true class label, the TSI relaxation reduces to minimizing the average negative log-probability that the observer chooses the correct class. 
The only difference from the binary case is that the observer now outputs an $L$-element probability vector, and $v[x](y)$ denotes the probability it assigns to class $y$. Consequently, the infimum in \eqref{eq:conditional-v-entropy} is evaluated exactly as in the binary case.
The ability to readily compute $\mathcal{V}$-info for multi-class problems represents one important advantage over ROC-based measures.

\subsection{Distinctions from signal detection measures}

{ 
$\mathcal{V}$-information and signal detection theory (SDT)–based measures
represent fundamentally different approaches to assessing task-based image
quality. As an example, consider a binary detection task.
The area under the receiver operating characteristic curve (AUC) 
has a well-known probabilistic
interpretation as:
\begin{equation}
\mathrm{AUC} = \Pr\big(s(\mathbf{X}^+) > s(\mathbf{X}^-)\big),
\label{auc-eq}
\end{equation}
where $s(\cdot)$ denotes an observer’s scalar response or score, and  $\mathbf{X}^+$ and
$\mathbf{X}^-$ denote image data corresponding to positive (signal present) and negative (signal absent) task outcomes,
respectively.
 The function $s(X)$ denotes the observer’s scalar decision variable (or test statistic) computed from the image, with larger values of $s(X)$ indicating stronger evidence in favor of signal presence. 
As such, AUC quantifies the probability that a randomly selected
positive case is ranked higher than a randomly selected negative case. This
interpretation makes AUC a measure of rank-based discriminability that depends
only on the relative ordering of observer responses and is invariant to any
strictly monotone transformation of the scores. 
 It is therefore insensitive to the numerical values of the observer’s
predictions, including how strongly they favor one outcome over another or how
much uncertainty they reflect.

In contrast, $\mathcal{V}$-info is an observer-aware, information-theoretic quantity that
measures the reduction in task-relevant uncertainty about the variable $\mathbf{Y}$
enabled by access to the image data $\mathbf{X}$ for observers belonging to a specified
class $\mathcal{V}$. Because it is defined using a principled probabilistic loss function,
$\mathcal{V}$-info rewards inferences whose reported uncertainty is well calibrated to the
observer’s posterior belief about the task outcome. In this setting, confidence does not
refer to a heuristic score, but to the extent to which the image strongly shifts the
observer’s posterior belief toward one task outcome relative to prior expectation,
reflecting the strength of the diagnostic evidence provided by the image. Accordingly,
$\mathcal{V}$-info characterizes both the diagnostic information supplied by the image and
the residual ambiguity that remains after observation, favoring images that support
reliable and unambiguous task inference over those that merely yield correct discrete
decisions.

Unlike AUC, which summarizes decision-level separability by evaluating how well cases associated with different task outcomes can be ranked across decision thresholds, $\mathcal{V}$-info quantifies the amount of task-relevant information accessible to an observer prior to committing to any specific decision rule, operating point, or error-cost structure. In this sense, AUC characterizes performance for a particular decision formulation, whereas $\mathcal{V}$-info addresses the more fundamental question of how informative an image is for the task from the observer’s perspective.

These conceptual differences have important practical implications. In regimes
where observer performance is far from saturation and discriminability is the
dominant limiting factor, signal detection theory-based measures and $\mathcal{V}$-info may
exhibit similar trends, reflecting consistent improvements in task-relevant
signal separation. However, as observer performance approaches saturation, or as
observer capacity becomes the primary constraint, detection-based metrics such as AUC, which are ranked-based,
can lose sensitivity and provide a limited ability to distinguish between imaging conditions.
In such cases, however, $\mathcal{V}$-info continues to evolve, capturing changes in posterior uncertainty, calibration,
and information accessibility that are invisible to rank-based decision metrics.
These behaviors are revealed in the numerical experiments presented below and
explained analytically in Appendix G.
}

\vspace{-0.1in}
\subsection{Computation of $\mathcal{V}$-information}

Consider a distribution of image data \( \mathbf{X} \) and a diagnostic task of predicting \( \mathbf{Y} \) from \( \mathbf{X} \) by use of a  parameterized NO model that can implement the family of mappings \( \mathcal{V} = \{v_\theta\}_{\theta \in \Theta} \), where \( \theta \in \Theta \) denotes the set of model parameters (e.g., neural network weights). 
 Specifically, the quantity \(v_\theta[x_i](y_i)\) denotes the predicted probability assigned by this NO to the true label \(y_i\) for input \(x_i\). 
Also consider that a finite dataset \( \{(x_i, y_i)\}_{i=1}^N \) of \( N \) image-label pairs sampled from the joint distribution \( P_{\mathbf{X}, \mathbf{Y}} \) are available. 
The entropy term $H(\mathbf{Y})$ in \eqref{ouinfo}  can be computed as:
\begin{equation}
H(\mathbf{Y})
\;=\;
- \sum_{y\in\mathbf{Y}}
      P_{\mathbf{Y}}(y)\,
      \log P_{\mathbf{Y}}(y),
\label{Hy}
\end{equation}

\noindent where $P_{\mathbf{Y}}(y)$ is the probability distribution of the label.
The conditional $\mathcal{V}$-entropy in \eqref{ouinfo}, $ H_{\mathcal{V}}[\mathbf{Y}|\mathbf{X}]$, can be computed as:
\begin{equation}
 H_{\mathcal{V}}[\mathbf{Y}|\mathbf{X}] = \arg\min_\theta\mathbb{E}_{x,y\sim P_{\mathbf{X},\mathbf{Y}}}[-\log v_\theta[x](y)],
 \label{opt-vinfo}
\end{equation}
which  corresponds to minimizing the cross-entropy loss function:
\vspace{-0.07in}
\begin{equation}
    L(\theta)=\frac{1}{N}\sum_{i=1}^N-\log v_\theta[x_i](y_i).
\end{equation}
Standard gradient-based optimization methods that are commonly used for machine learning applications\cite{kingma2014adam} can be employed for this purpose.

{ 
In practical implementations, predictive $\mathcal{V}$-info does not
require a separate optimization procedure beyond that already performed when
training the observer model itself. When the observer class $\mathcal{V}$ is
instantiated using standard learning paradigms, $\mathcal{V}$-info is
naturally obtained through conventional training objectives, such as
cross-entropy minimization for classifiers or conditional likelihood training
for generative models. Once training is complete, estimating
$\mathcal{V}$-info reduces to evaluating cross-entropy or log-likelihood
terms on held-out data, which is computationally comparable to standard model
evaluation.

As a result, the computational cost of estimating $\mathcal{V}$-info
scales primarily with the cost of evaluating the trained observer model, rather
than directly with the dimensionality of the image data. In this sense,
$\mathcal{V}$-info introduces minimal additional computational overhead
relative to existing task-based evaluation pipelines, making it practical for
use in large-scale imaging studies and system comparisons.
}

\section{Description of Case Studies}
\label{application}

Studies were conducted to demonstrate the practical utility of $\mathcal{V}$-info as a task-based  IQ metric. A deep neural network (DNN)-based image restoration method was evaluated with consideration of signal known statistically/background known statistically (SKS/BKS) signal detection tasks. 
$\mathcal{V}$-info and traditional signal detection measures were employed to quantify signal detection performance  with consideration of sub-ideal DNN-based NOs. 
The relative behavior of these metrics was  studied to corroborate the usefulness of $\mathcal{V}$-info for objectively assessing IQ.

\vspace{-0.1in}
\subsection{Data Preparation}
\label{data-preparation}

Structural brain MRI data from the Human Connectome Project (HCP) Young Adult dataset~\cite{hcp1200} were employed in virtual imaging studies. This dataset comprises three-dimensional (3D) T1-weighted MRI volumes acquired at 3 Tesla (high-field MRI) from 1,113 healthy young adults. Each 3D volume was of dimension  $260\times311\times260$ voxels. Ten central slices containing both white matter and gray matter were extracted from each 3D MRI volume. These selected 11,130 slices were  padded to a final dimension of $288\times320$ pixels.

\subsubsection{Signal-present and signal-absent image generation}

The extracted two-dimensional (2D) MRI slices were directly used as signal-absent (background) objects  $\mathbf{f}_\mathbf{b}$. To generate corresponding signal-present objects, synthetic Gaussian signals mimicking lesions were computationally inserted into these background objects. 

Each  signal $\mathbf{f}_\mathbf{s}(x,y)$ was defined as a Gaussian function:
\vspace{-0.05in}
\begin{equation}
    \mathbf{f}_s(x,y)
    = \lambda A_s \exp\left(-\frac{(x - x_0)^2 + (y - y_0)^2}{2\sigma^2}\right),
\end{equation}
\noindent where \(A_s\) denotes the signal amplitude, \((x_0, y_0)\) indicates the location of the signal center that was assumed uniformly random in white matter regions, and \(\sigma=3\) pixels defined the spatial spread of the signal. The indicator function \(\lambda\equiv\mathbb{I}\left(\sqrt{(x - x_0)^2 + (y - y_0)^2} \leq 3\sigma\right)\) defined the
support of the signal to be a circular region of radius \(3\sigma\). Signal-present images, denoted by \(\mathbf{f}_{\mathbf{b+s}}\), were formed as:
\begin{equation}
    \mathbf{f}_{\mathbf{b+s}}(x,y) = \mathbf{f}_\mathbf{b}(x,y) + \mathbf{f}_\mathbf{s}(x,y).
\end{equation}

Objects with one signal inserted were utilized in binary signal detection tasks, while objects containing two signals (with distinct spatial locations) were prepared for three-class classification tasks.

\subsubsection{Low-field MRI data simulation}

The signal-present and signal-absent objects were virtually imaged to emulate stylized low-field MRI conditions. The low-field MRI simulation involved two main steps: (1) generation of noisy \textit{k}-space (frequency-domain) data, followed by (2) image reconstruction in the spatial domain~\cite{de2022deep}.

Initially, each spatial-domain object \(\mathbf{f}\) (either \(\mathbf{f}_\mathbf{b}\) or \(\mathbf{f}_{\mathbf{b+s}}\)) was transformed into \textit{k}-space using the two-dimensional discrete Fourier transform (DFT). To simulate the characteristic reduced spatial resolution of low-field MRI, a rectangular frequency mask of dimension $144\times160$ pixels was applied, preserving only the central low-frequency components. Independent and identically distributed zero-mean complex Gaussian noise was then added to these masked frequency components. After noise addition, the masked \textit{k}-space data were zero padded to match the numerical observer (neural network) input dimensions.


\vspace{-0.1in}

\subsection{Image Restoration Methods}
A supervised MRI image restoration problem was considered, where the goal was to estimate a  high-resolution
 image \( \mathbf{f} \) from a corresponding low-resolution image  \( \tilde{\mathbf{f}} \).
 Motivated by a low-field MRI restoration problem\cite{de2022deep}, these images will be referred to as high-field and low-field images, respectively.

\begin{figure}[!h]
    \centering
\includegraphics[width=0.95\linewidth]{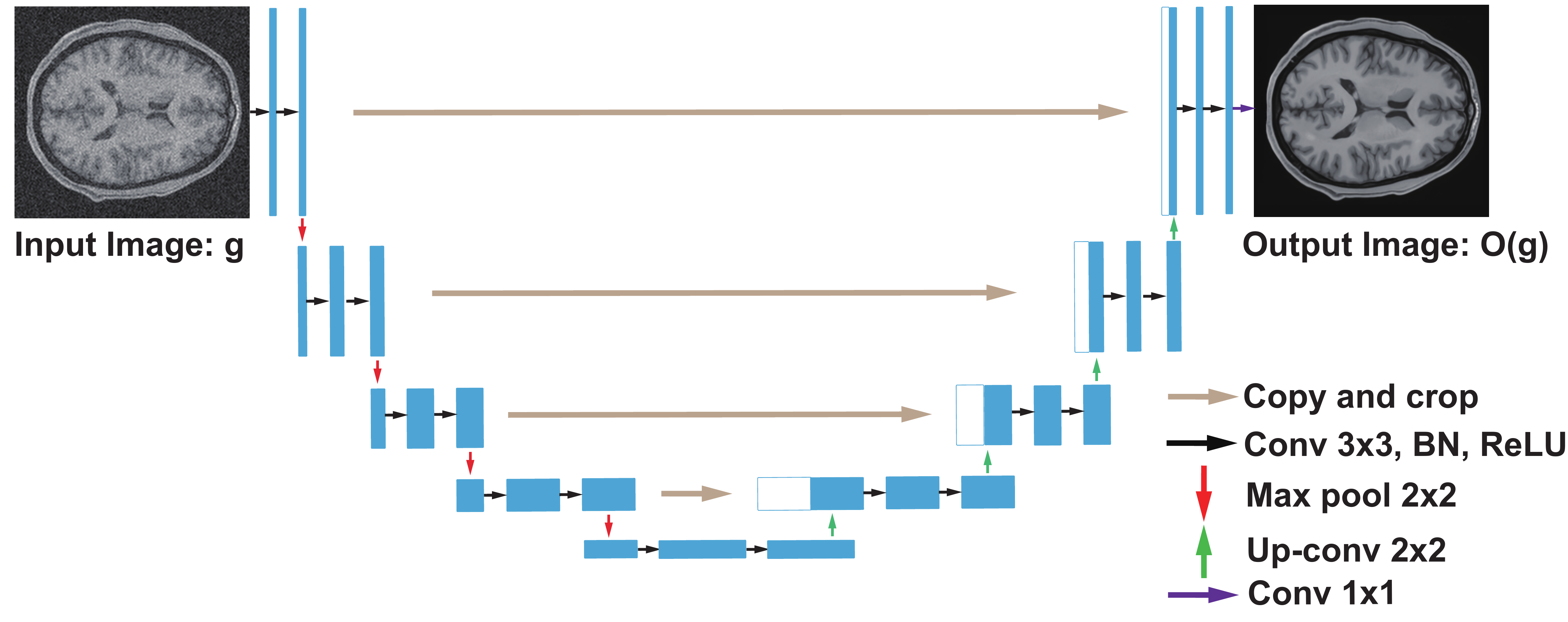}
    \caption{MRI restoration network architecture. The dimensions of the input and output images are 288 $\times$ 320.}
    \label{unet}
\end{figure}

The restoration process is denoted as:
\begin{equation}
    \hat{\mathbf{f}} = \mathcal{O}(\tilde{\mathbf{f}}; \theta),
\end{equation}
\noindent where $\tilde{\textbf{f}}$ is the low-field MRI image, \(\mathcal{O}\) denotes the  restoration operator parameterized by \(\theta\), and $\hat{\textbf{f}}$ denotes the  estimate of the high-field image.
A  U-Net-based architecture~\cite{ronneberger2015u}, illustrated in Fig.~\ref{unet}, was employed
to implement the restoration operator.
Given a collection of paired data  $\left\{\tilde{f_i},f_i\right\}_{i=1}^N$, where $\tilde{f_i}$ and $f_i$ denote the low-field and high-field target  images, respectively, the restoration network was trained using a mean-squared-error (MSE) loss function:
\vspace{-0.1in}
\begin{equation}
    \mathcal{L}_{\mathrm{MSE}}(\theta) = \frac{1}{N} \sum_{i=1}^N \|\mathcal{O}(\tilde{f_i}; \theta) - f_i\|_2^2.
    \vspace{-0.05in}
\label{mse}
\end{equation}
\noindent The network architecture and training details are described in Appendices~\ref{restoration-details} and \ref{Restoration-training}.

\vspace{-0.15in}
\subsection{Objective IQ Assessments}

\subsubsection{Detection and classification tasks}
Two task settings were investigated: a binary signal detection task and a three-class signal classification task. In the binary detection task, the goal was to determine the presence or absence of a signal. In the three-class classification task, the objective was to distinguish among signal-absent, one-signal-present, and two-signals-present conditions. Several signal and background configurations were varied to study how $\mathcal{V}$-info responds across different tasks. Additional details are provided in Appendix~\ref{signal-setting}.

\subsubsection{DNN-based NOs}
Sub-ideal CNN- and ResNet-based NOs were utilized to perform the signal detection and classification tasks. CNN-based NOs were used for the binary detection task, while ResNet-based NOs were used for the three-class classification task. To study the effect of observer capacity, each NO was instantiated with multiple configurations by varying the network depth (i.e., the number of convolutional or residual layers). Implementation details of the NO architectures are provided in Appendix~\ref{NO-setting}. 

\subsubsection{$\mathcal{V}$-info and traditional task-based IQ metrics} 
\label{NO-Computation}
The quality of low-field, estimated high-field, and ground truth high-field images was objectively assessed by computing $\mathcal{V}$-info according to 
(\ref{ouinfo}).
For comparison, traditional task-based metrics were also computed.
For the binary signal detection task,  AUC was computed.
Since ROC analysis is not readily applicable to multi‐class problems, classification accuracy was used to evaluate NO performance for the three‐class signal classification task.
This was defined as the proportion of image samples for which the predicted class matched the true class label.
NO performance is examined for both balanced and imbalanced cases.
During training, \( \mathcal{V} \)-info was computed on the training set described below by minimizing the average negative log-likelihood of the observer’s predictions, as described in \eqref{opt-vinfo}, while AUC and accuracy were measured on the test set. AUC was computed based on the observer’s predicted confidence scores (i.e., softmax outputs) for the target class, using the \texttt{scikit-learn} implementation to estimate AUC~\cite{scikit-learn}.

\subsubsection{Dataset configuration}
For the binary detection task, the training set included 8,904 signal-present and 8,904 signal-absent images, with 1,113 images per group in the test set. For the three-class classification task, the training set comprised 8,904 images each from signal-absent, one-signal-present, and two-signals-present groups, with 1,113 test images per group. 
In the imbalanced data setting, the binary classification task used 8,904 signal-absent and 1,781 signal-present training images, while the three-class task used 8,904 signal-absent, 890 one-signal-present, and 890 two-signals-present training images. All reported results were obtained by averaging over five independent runs. The error bars were computed using the standard deviation across these runs.

\vspace{-0.15in}
\section{Results of Case Studies}
\label{results}

\begin{figure}[!h]
    \centering
    \begin{subfigure}{0.32\linewidth}
        \includegraphics[width=\linewidth]{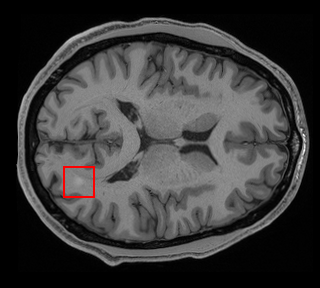}
    \end{subfigure}
    \begin{subfigure}{0.32\linewidth}
        \includegraphics[width=\linewidth]{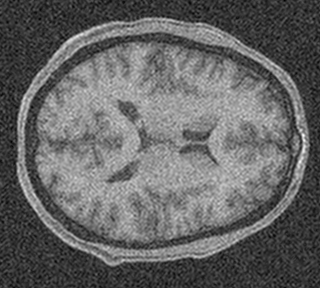}
    \end{subfigure}
    \begin{subfigure}{0.32\linewidth}
        \includegraphics[width=\linewidth]{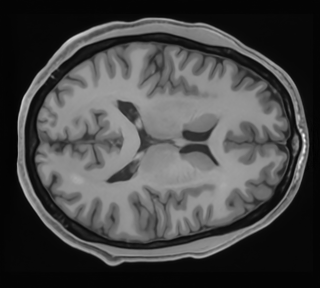}
    \end{subfigure}
    
    \begin{subfigure}{0.32\linewidth}
        \includegraphics[width=\linewidth]{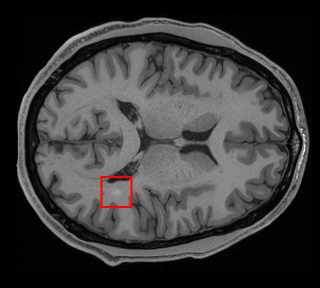}
    \end{subfigure}
    \begin{subfigure}{0.32\linewidth}
        \includegraphics[width=\linewidth]{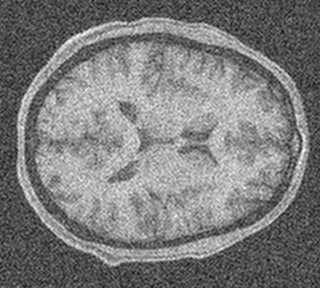}
    \end{subfigure}
    \begin{subfigure}{0.32\linewidth}
        \includegraphics[width=\linewidth]{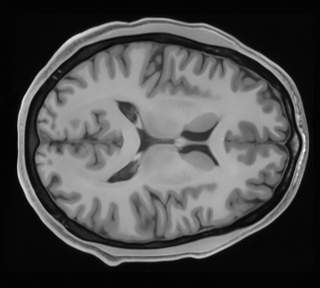}
    \end{subfigure}

    \begin{subfigure}{0.32\linewidth}
        \includegraphics[width=\linewidth]{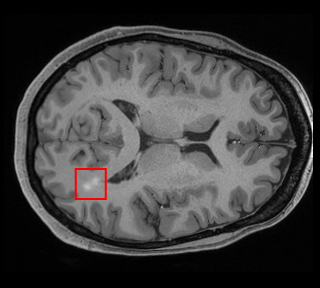}
    \end{subfigure}
    \begin{subfigure}{0.32\linewidth}
        \includegraphics[width=\linewidth]{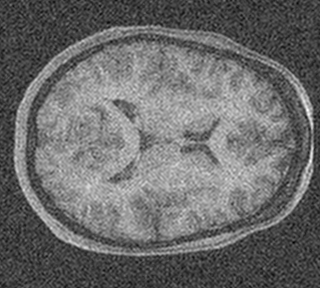}
    \end{subfigure}
    \begin{subfigure}{0.32\linewidth}
        \includegraphics[width=\linewidth]{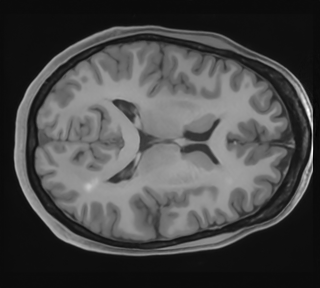}
    \end{subfigure}
    
    \begin{subfigure}{0.32\linewidth}
        \includegraphics[width=\linewidth]{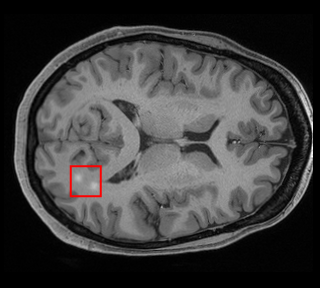}
        \caption{}
    \end{subfigure}
    \begin{subfigure}{0.32\linewidth}
        \includegraphics[width=\linewidth]{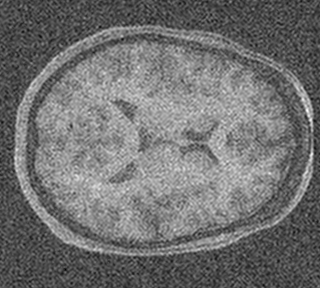}
        \caption{}
    \end{subfigure}
    \begin{subfigure}{0.32\linewidth}
        \includegraphics[width=\linewidth]{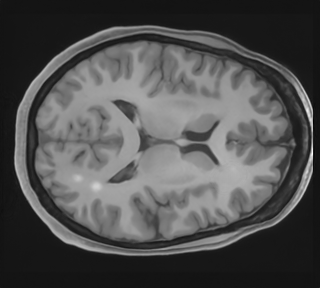}
        \caption{}
    \end{subfigure}
    \caption{Representative examples of high-field, low-field, and restored high-field MRI images are shown. 
    Columns correspond to (a) high-field MRI, (b) low-field MRI, and (c) restored high-field MRI.
    The red box highlights the signal.
    The first and second rows show one-signal-present images with background noise levels of 35 and 45, respectively, and a signal amplitude of 0.1. The third and fourth rows show two-signal-present images with noise levels of 35 and 45 and signal amplitude of 0.15, respectively.}
    \label{restoration visual}
\end{figure}

\subsection{Preliminary IQ Assessments}

Representative examples of low-field, estimated high-field, and ground truth high-field MRI are shown in Fig.~\ref{restoration visual}.
As compared to the original low-field image, the noise in the restored image was reduced and the boundary between white and gray matter appeared subjectively clearer. 
However, other high-frequency components that include textures and features within the white matter, which were present in high-field MRI images, were not reliably recovered. 

\begin{table}[h]
\caption{IQ comparison between low-field and restored high-field MRI images in terms of SSIM and PSNR.}
\centering
\begin{tabular}{cccc}
\toprule
 &  \begin{tabular}[c]{@{}c@{}}Image\\Quality Metric\end{tabular} & \begin{tabular}[c]{@{}c@{}}Background\\ Noise Level: 35\end{tabular} & \begin{tabular}[c]{@{}c@{}}Background\\ Noise Level: 45\end{tabular} \\ \hline
\multirow{2}{*}{\begin{tabular}[c]{@{}c@{}}Low-field\\ MRI\end{tabular}}        & SSIM & 0.432$\pm$0.0020 & 0.383$\pm$0.0032 \\
& PSNR & 15.682$\pm$0.222 & 14.438$\pm$0.370 \\ \hline
\multirow{2}{*}{\begin{tabular}[c]{@{}c@{}}Estimated\\ High-field MRI\end{tabular}} & SSIM & 0.893$\pm$0.0031 & 0.884$\pm$0.0043 \\
& PSNR & 30.047$\pm$0.147 & 29.312$\pm$0.192 \\ \hline
\end{tabular}
\label{iq2}
\end{table}

Structural similarity index measure (SSIM) and peak signal-to-noise ratio (PSNR) values are provided in Table~\ref{iq2}. These data are consistent with the subjective visual assessments and confirm that the image restoration operation yielded improvements in these conventional IQ metrics.

\begin{figure}[!htb]
    \centering
    \begin{subfigure}{0.45\linewidth}
        \includegraphics[width=\linewidth]{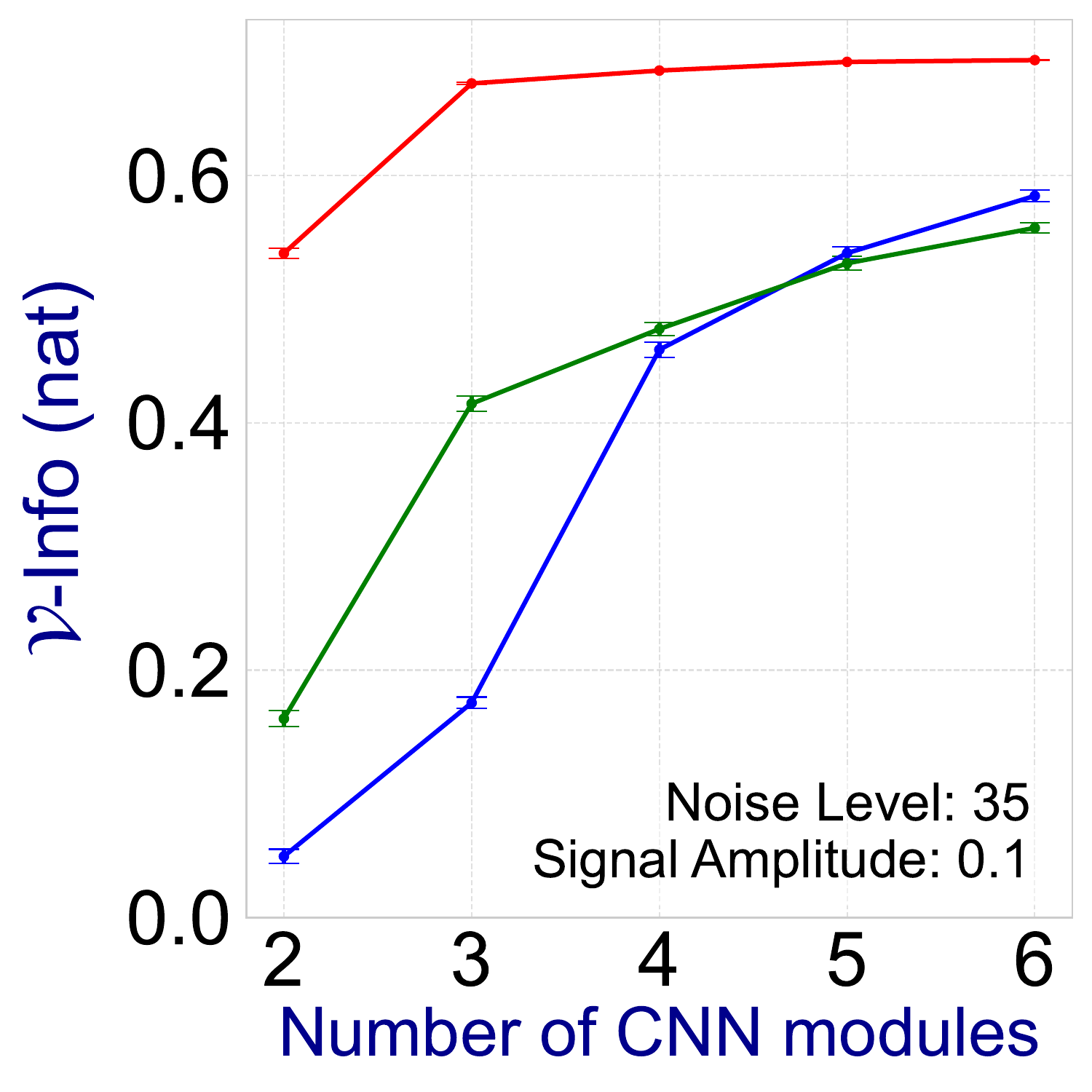}
    \end{subfigure}
        \hspace{0.1in}
     \begin{subfigure}{0.45\linewidth}
        \includegraphics[width=\linewidth]{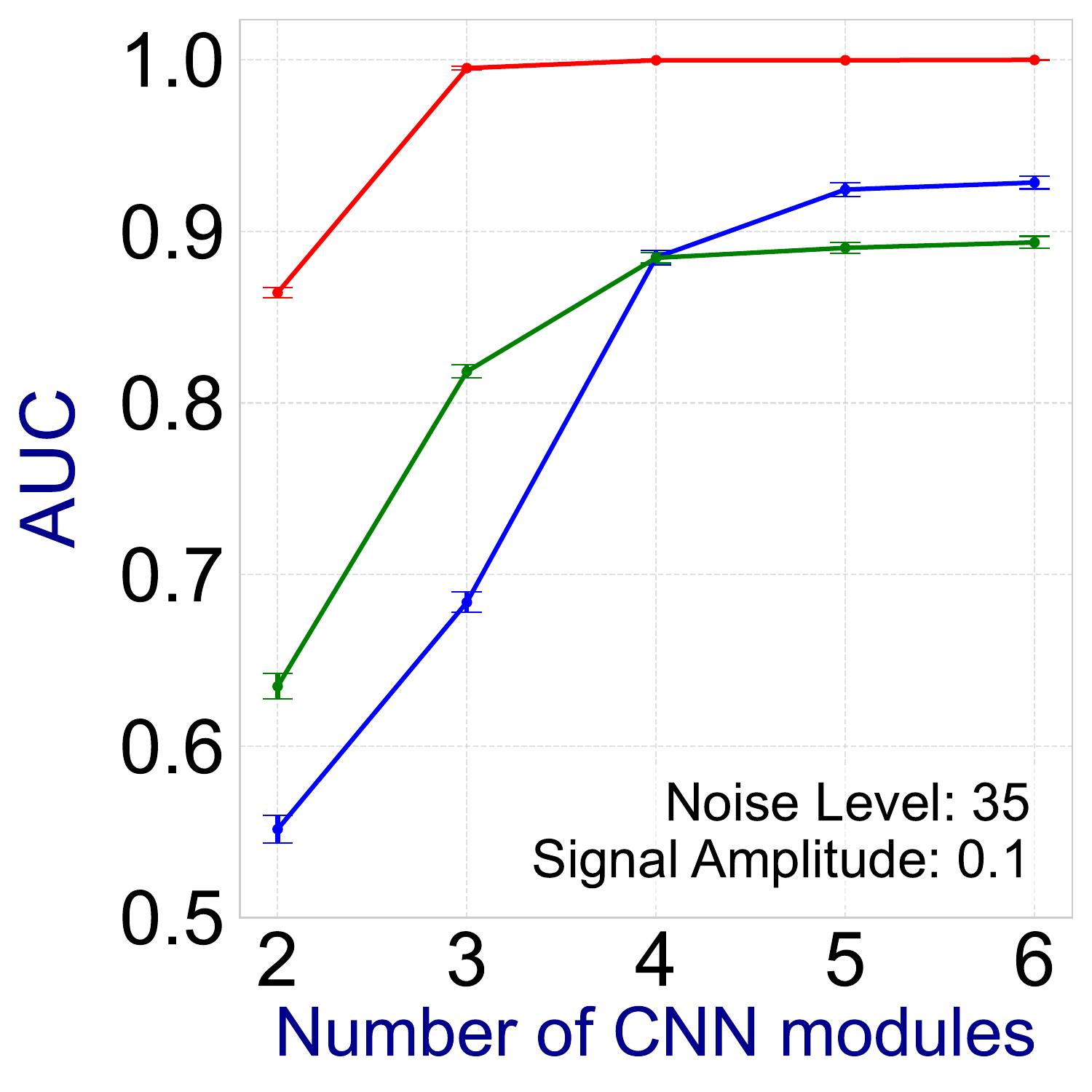}
    \end{subfigure}

     \begin{subfigure}{0.45\linewidth}
        \includegraphics[width=\linewidth]{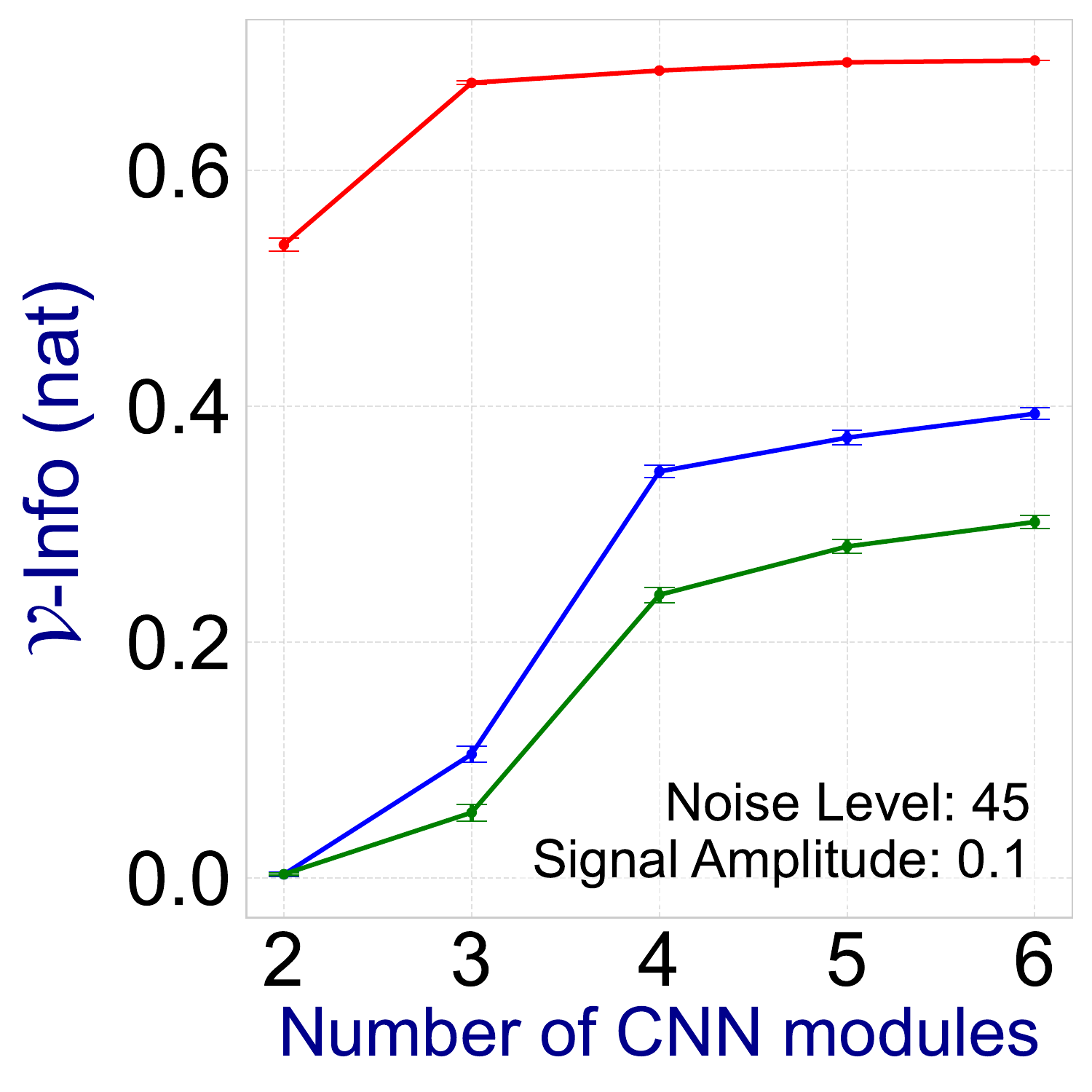}
    \end{subfigure}
        \hspace{0.1in}
    \begin{subfigure}{0.45\linewidth}
        \includegraphics[width=\linewidth]{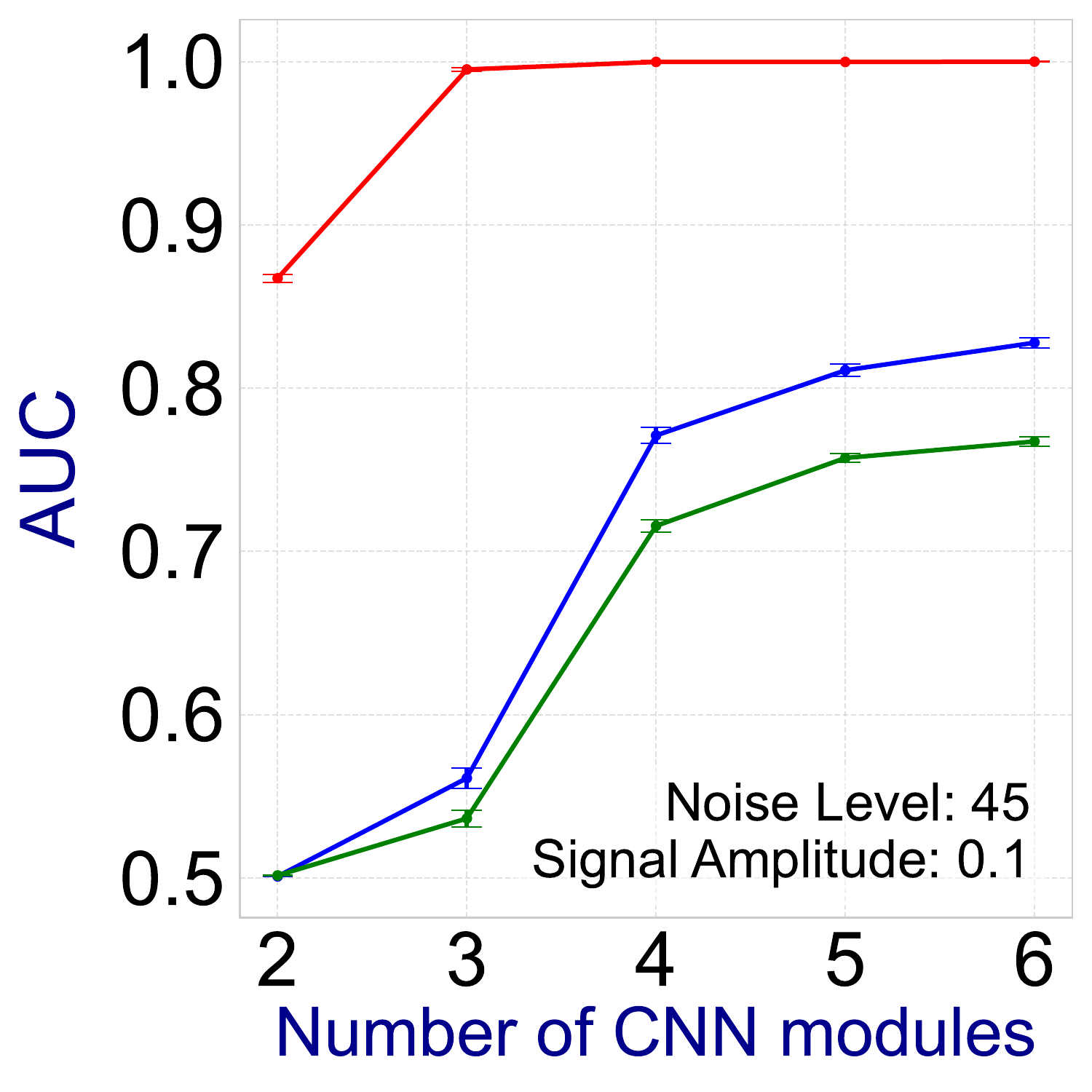}
    \end{subfigure}
 
    \begin{subfigure}{0.45\linewidth}
        \includegraphics[width=\linewidth]{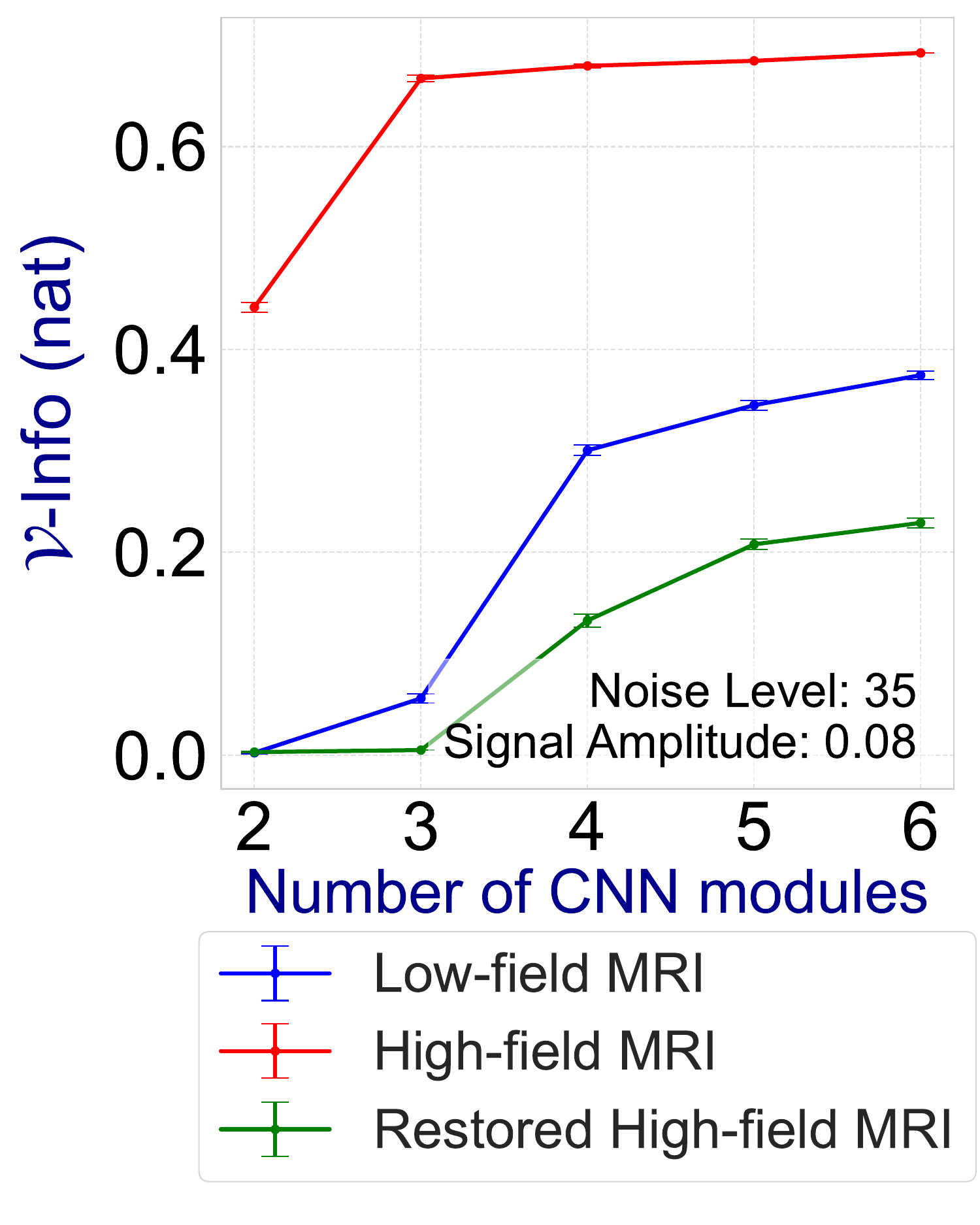}
        \caption{}
    \end{subfigure}
        \hspace{0.1in}
    \begin{subfigure}{0.45\linewidth}
        \includegraphics[width=\linewidth]{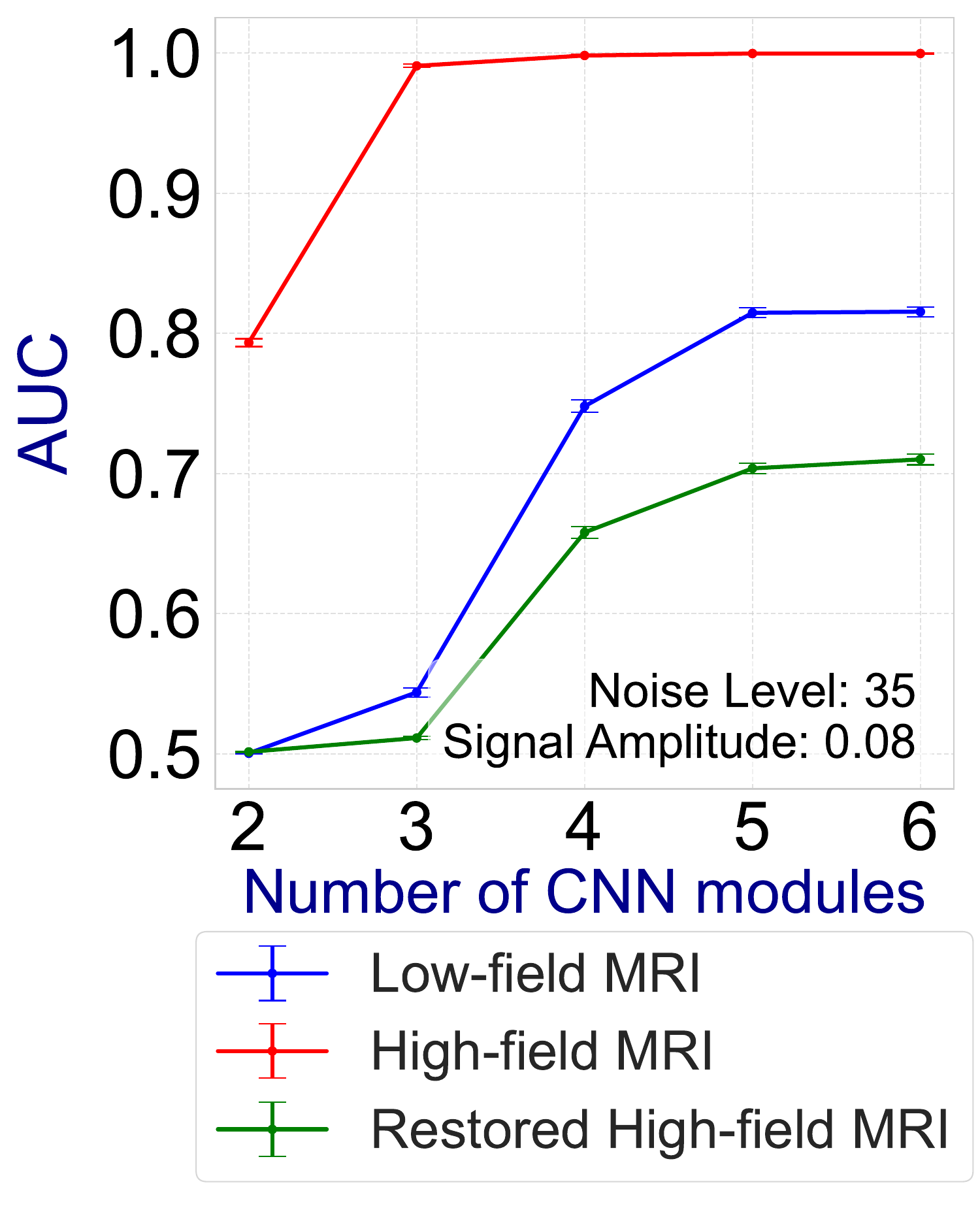}
        \caption{}
    \end{subfigure}
     \vspace{-0.05in}
     \caption{Observer performance on the binary signal detection task using a balanced dataset as quantified by $\mathcal{V}$-info (a) and AUC (b), shown as functions of the number of CNN NO modules in the observer. 
    Both metrics exhibit similar overall trends, indicating improved performance with increasing observer capacity across low-field, restored, and high-field image types. 
    However, as shown in the first and third rows, while AUC saturates at higher capacities and fails to distinguish further performance gains, $\mathcal{V}$-info continues to increase, providing a more sensitive measure of observer performance in these regimes.}
    \label{v-info and auc}
\end{figure}


\subsection{Task-Based IQ Assessments}
\subsubsection{Binary signal detection tasks (balanced data)}

\begin{figure}[!h]
\vspace{-0.05in}
  \centering
  \begin{subfigure}{0.32\linewidth}
    \includegraphics[width=\linewidth]{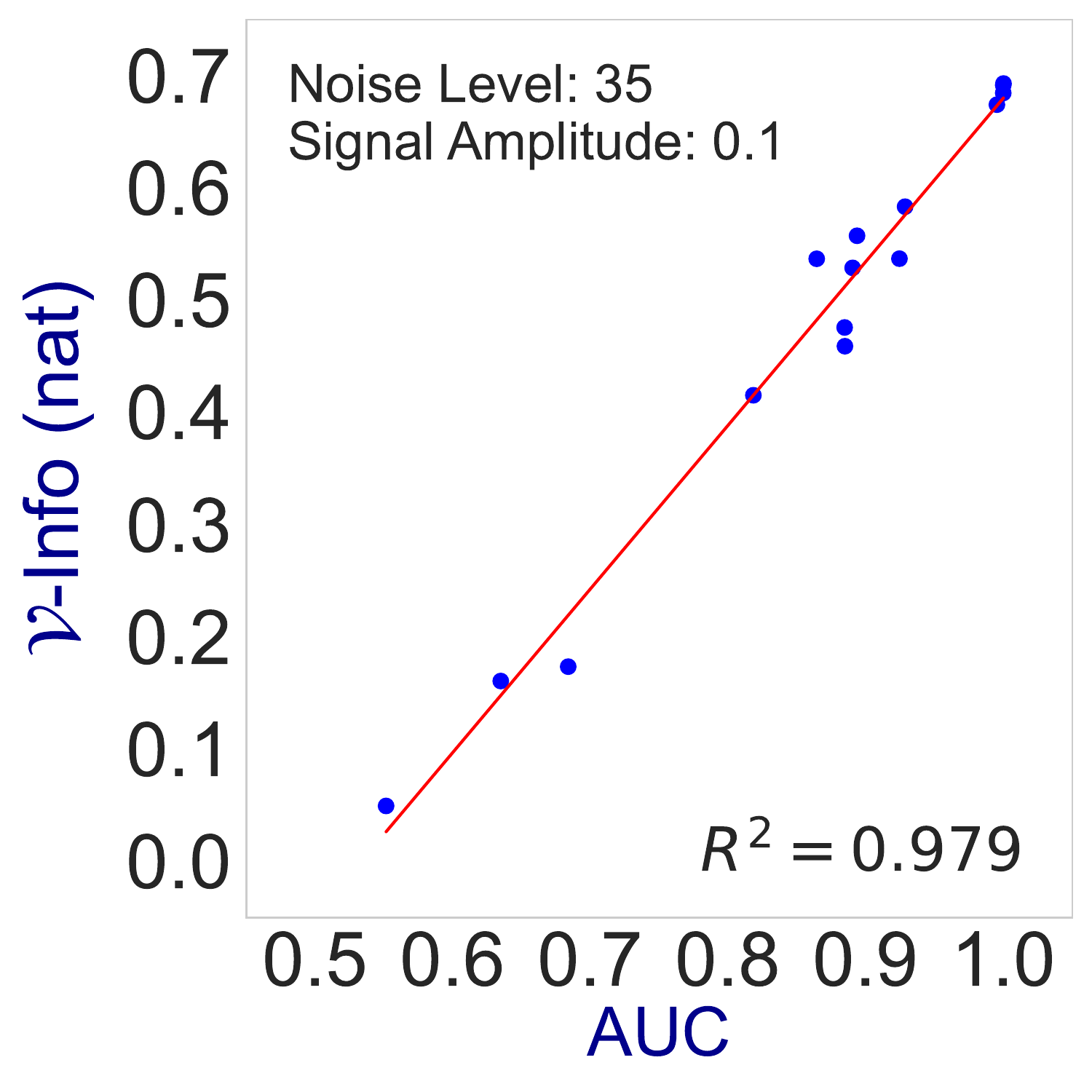}
    \caption{Low-field MRI}
  \end{subfigure}\hfill
  \begin{subfigure}{0.32\linewidth}
    \includegraphics[width=\linewidth]{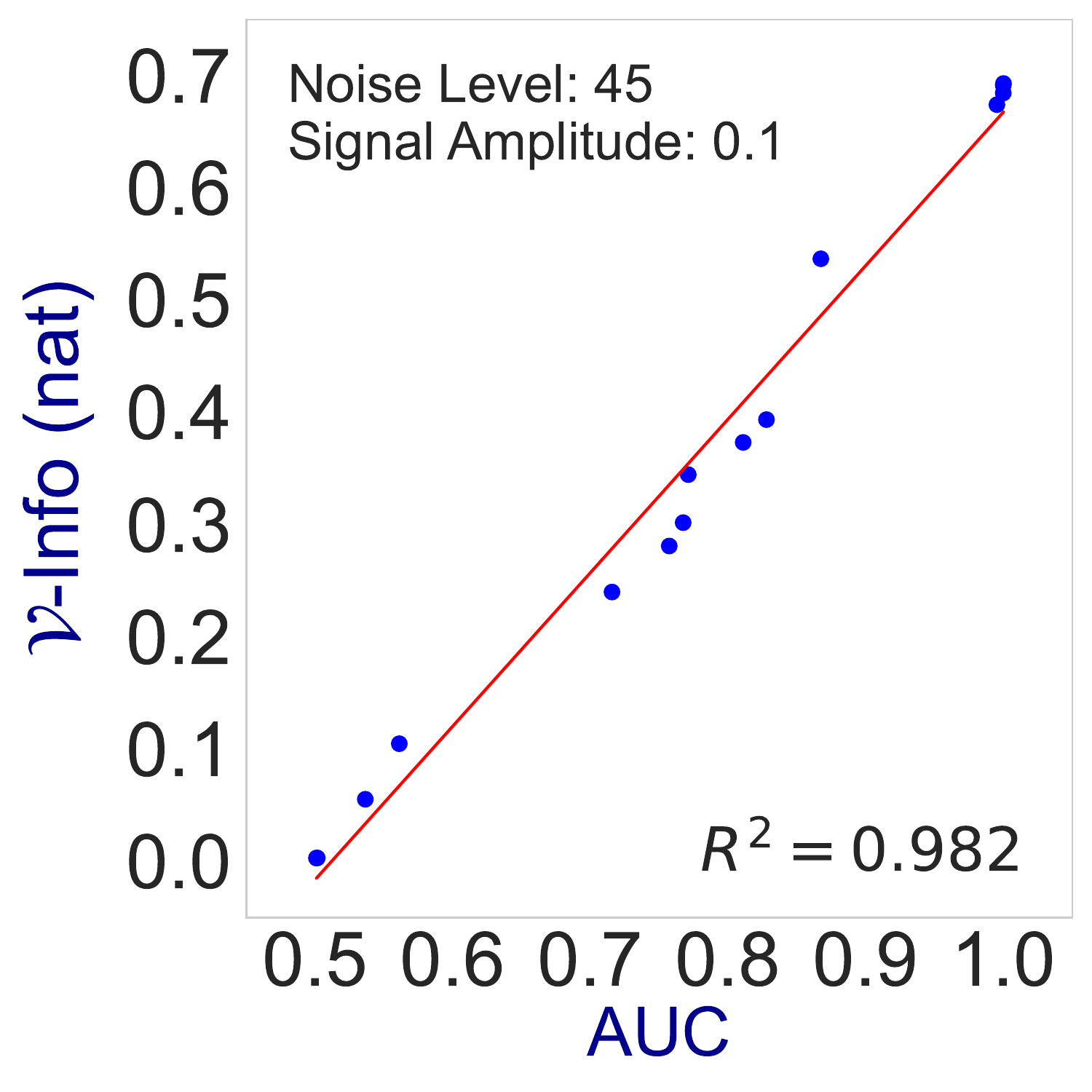}
    \caption{High-field MRI}
  \end{subfigure}\hfill
  \begin{subfigure}{0.32\linewidth}
    \includegraphics[width=\linewidth]{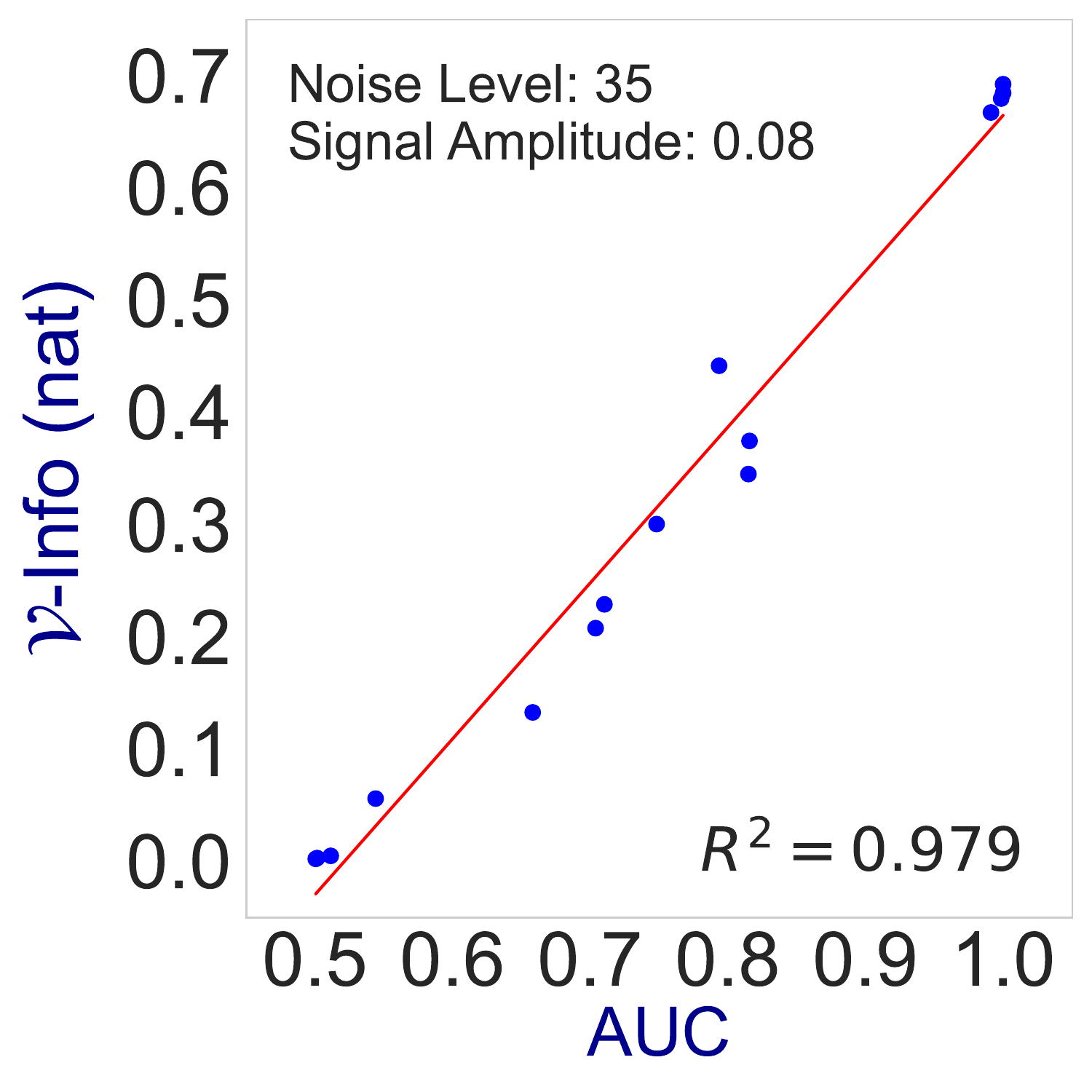}
    \caption{Restored MRI}
  \end{subfigure}
  \caption{Relationship between $\mathcal{V}$-info and AUC for CNN-based observers on the binary signal-detection task using a balanced dataset.
  Blue dots denote observer performance achieved with five capacity levels (2 – 6 CNN NO modules) under each imaging condition (low-field, high-field, restored).  
  A clear linear dependence is observed, with the coefficient of determination $R^{2}$ approaching 1.}
  \label{vinfo_auc_cnn}
  \vspace{-0.1in}
\end{figure}

The results for the binary signal detection tasks using a balanced dataset are described here.
 Figure~\ref{v-info and auc} presents observer performance quantified by $\mathcal{V}$-info and AUC across three background and signal settings (see Appendix~\ref{signal-setting}) and five levels of CNN-based NO capacity.
Both $\mathcal{V}$-info and AUC consistently improved with increased NO capacity across all three background and signal settings.
For the case where the background noise level was 35 and the signal amplitude was 0.1, the NO performance on the restored images based on both $\mathcal{V}$-info and AUC exceeded the performance on the original degraded images at lower observer capacities ($<4$ CNN NO modules). However, for higher NO capacities, signal detection performance  on the original degraded images outperformed that on restored images,
as measured by both  $\mathcal{V}$-info and AUC. 
As shown in Fig.~\ref{vinfo_auc_cnn}, the relationship between $\mathcal{V}$-info and AUC remained highly linear ($R^2 \approx 0.99$) across multiple signal and background conditions.

While both $\mathcal{V}$-info and AUC consistently increased with increased observer capacity,
their behaviors differed  at higher capacities. As shown in Fig.~\ref{v-info and auc}, AUC tended to saturate as observer capacity increased (e.g., from 5 to 6 CNN NO modules), while $\mathcal{V}$-info continued to increase. This indicates that $\mathcal{V}$-info provides a more sensitive measure of observer performance, particularly in regimes where AUC becomes less discriminative between conditions. 

\begin{figure}[!h]
    \centering
    \begin{subfigure}{0.4\linewidth}
        \includegraphics[width=\linewidth]{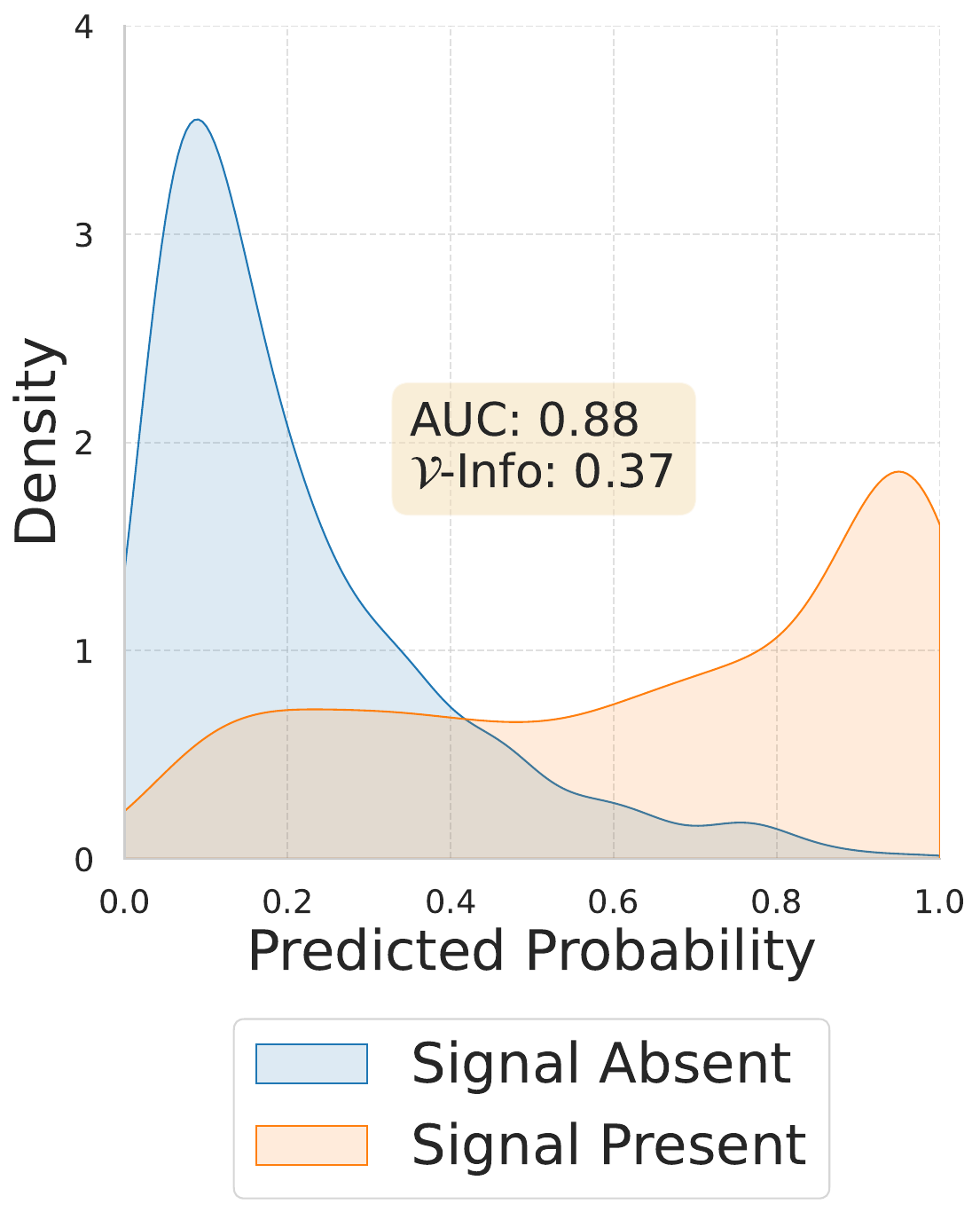}
    \end{subfigure}
     \hspace{0.1in}
    \begin{subfigure}{0.4\linewidth}
        \includegraphics[width=\linewidth]{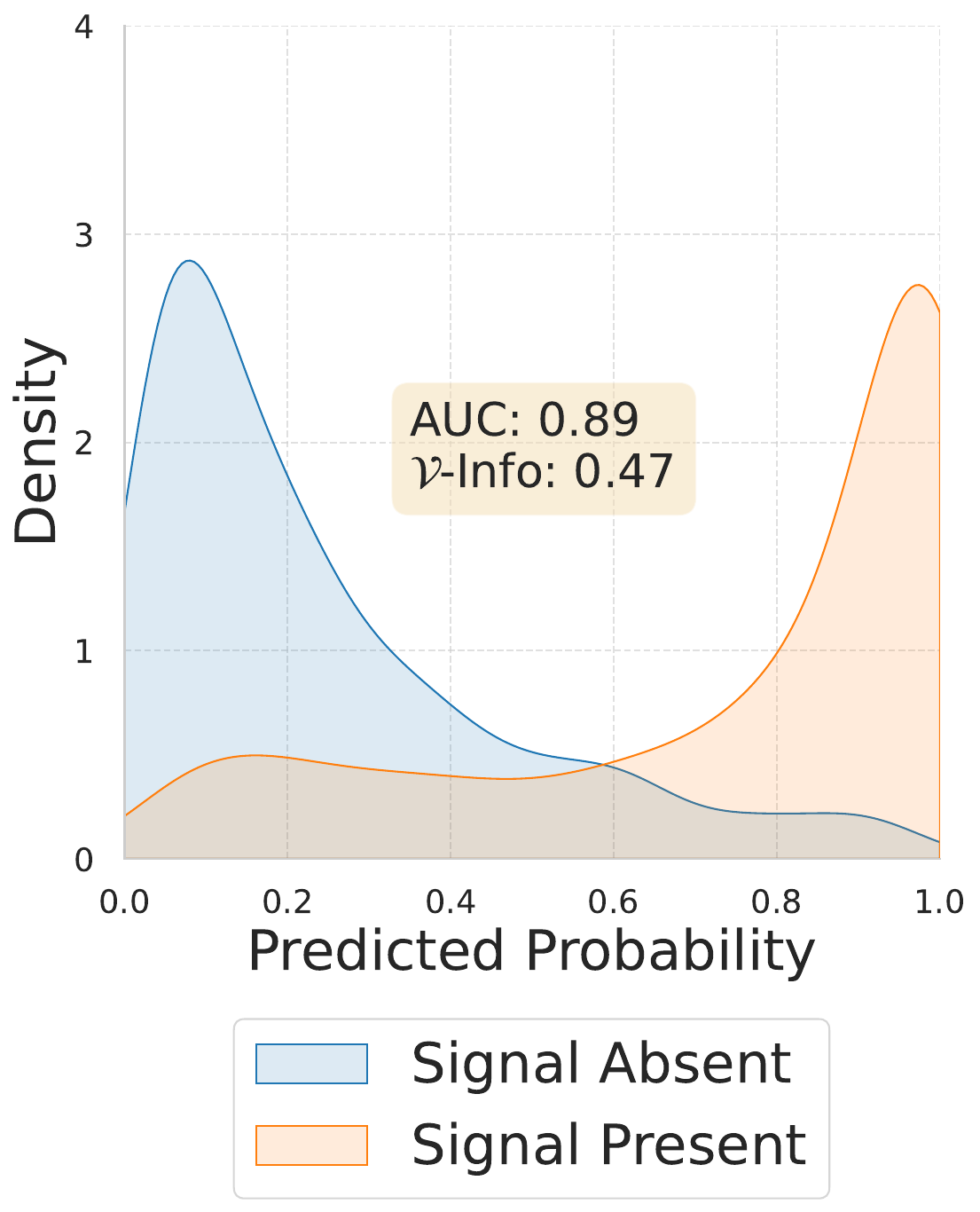}
    \end{subfigure}
    \caption{
Comparison of AUC and $\mathcal{V}$-info for a binary classification task. Blue and orange curves represent predicted probability distributions for signal-absent and signal-present classes, respectively. Although AUC values appear similar across the two examples, $\mathcal{V}$-info highlights a larger difference between the underlying probability distributions, illustrating its higher sensitivity.
}
    \label{binary plot}
      \vspace{-0.1in}
\end{figure}

Figure~\ref{binary plot} illustrates this point through a stylized example in which two predicted probability distributions with similar AUC values differ substantially in $\mathcal{V}$-info. This demonstrates that $\mathcal{V}$-info is capable of capturing differences in predictive confidence that AUC cannot easily distinguish, making it a valuable tool for measuring nuanced variations in observer performance.


\begin{figure}[!h]
    \centering
    \begin{subfigure}{0.44\linewidth}
        \includegraphics[width=\linewidth]{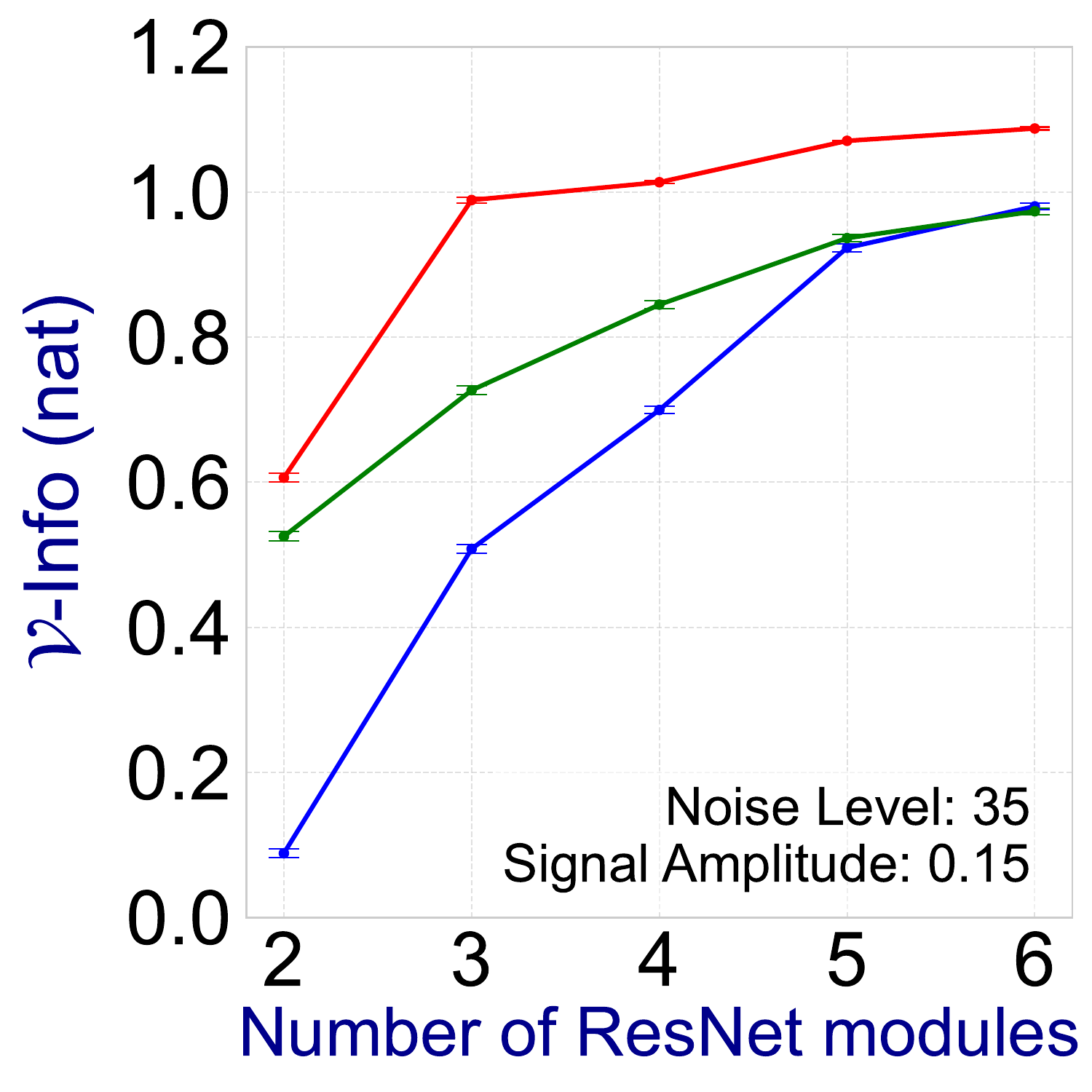}
    \end{subfigure}
        \hspace{0.1in}
    \begin{subfigure}{0.44\linewidth}
        \includegraphics[width=\linewidth]{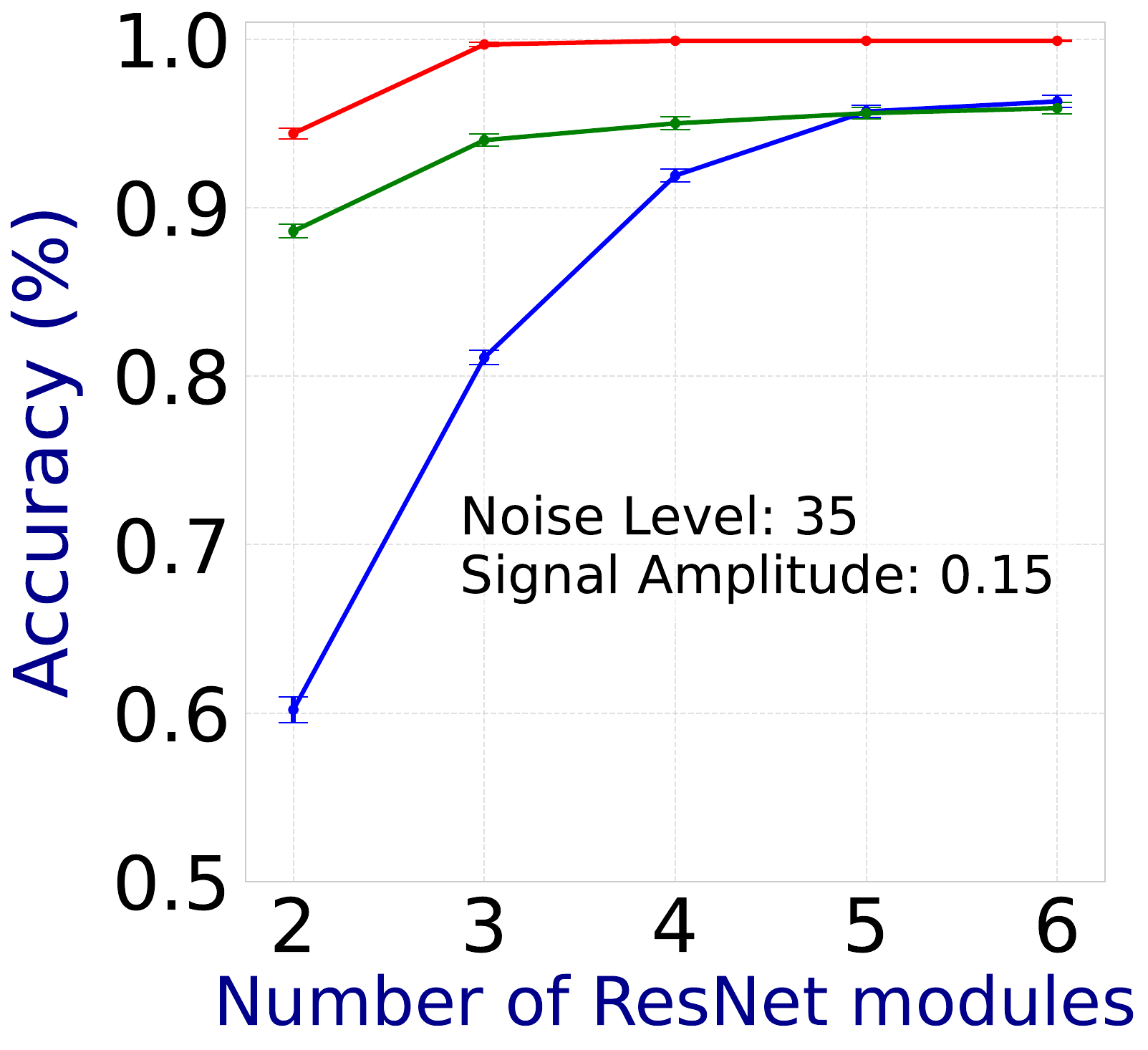}
    \end{subfigure}
    \begin{subfigure}{0.44\linewidth}
        \includegraphics[width=\linewidth]{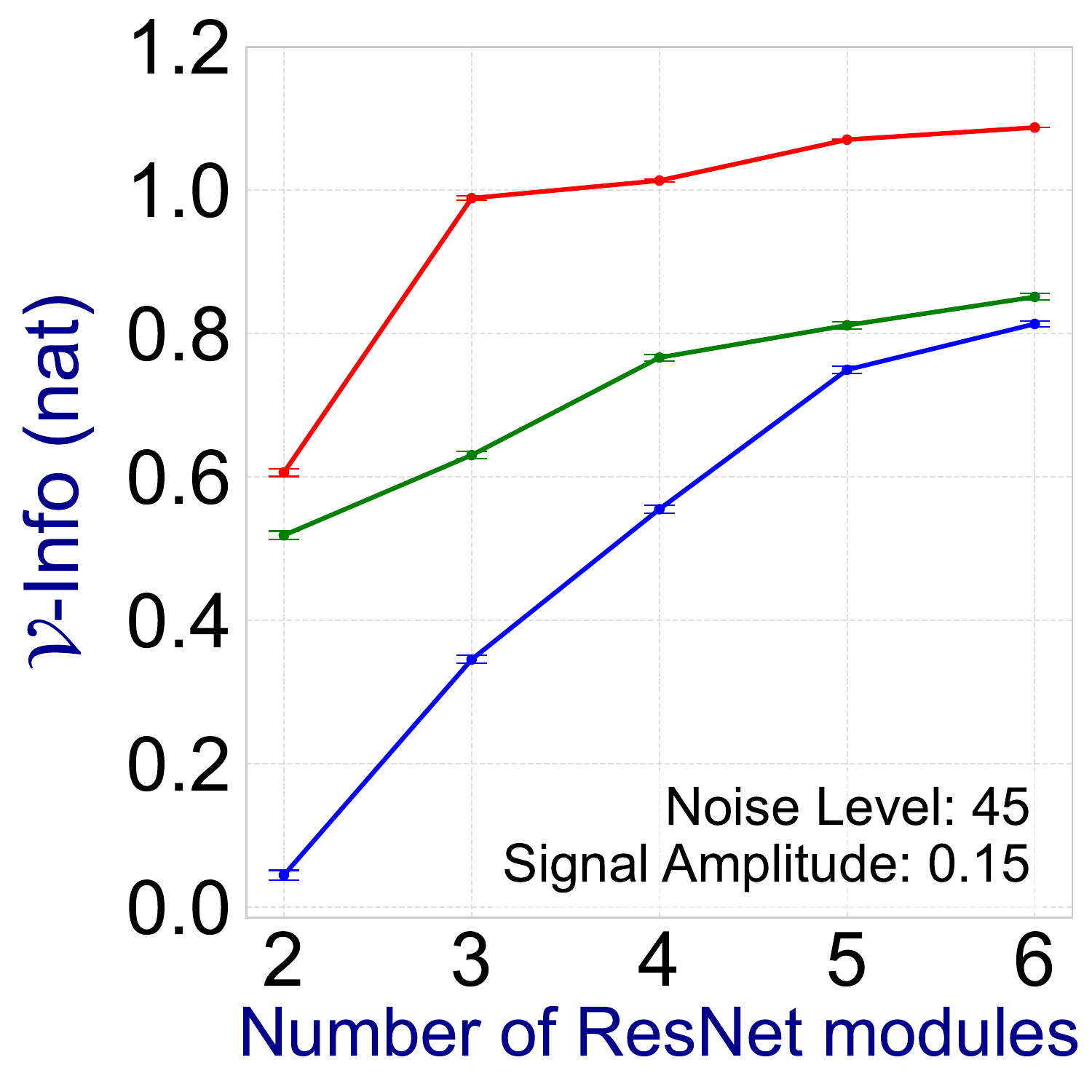}
    \end{subfigure}
         \hspace{0.1in}
    \begin{subfigure}{0.44\linewidth}
        \includegraphics[width=\linewidth]{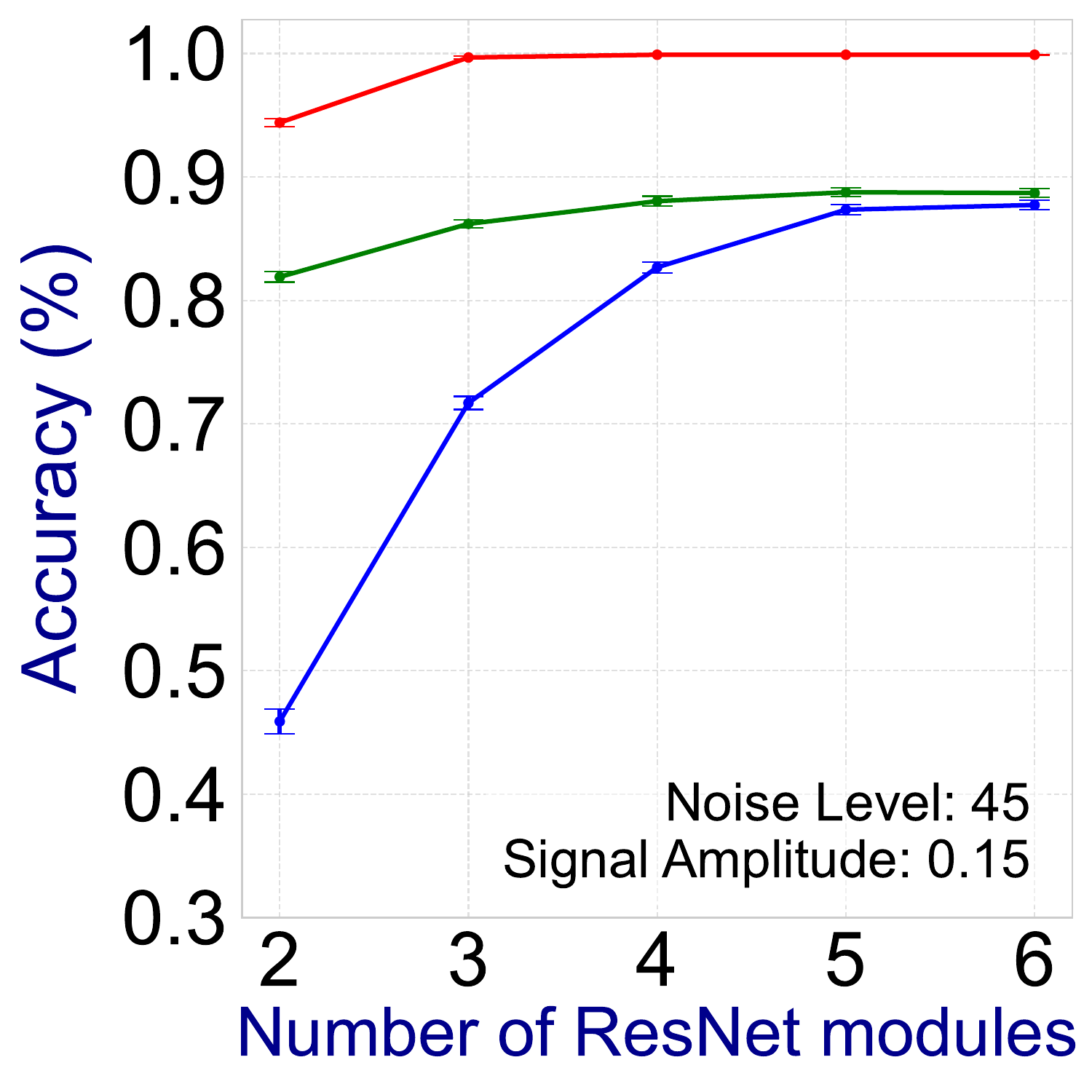}
    \end{subfigure}
    \begin{subfigure}{0.44\linewidth}
        \includegraphics[width=\linewidth]{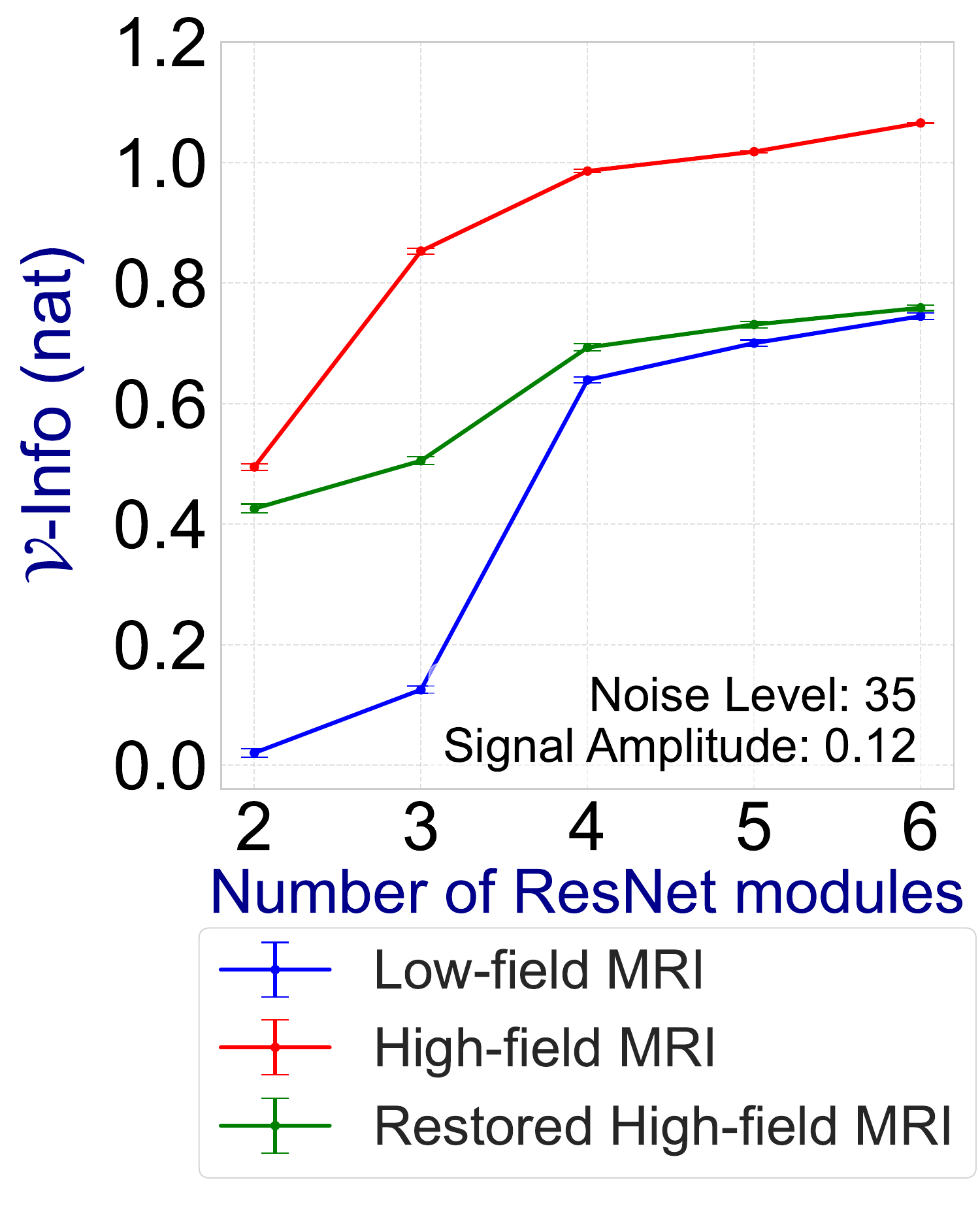}
        \caption{}
    \end{subfigure} 
        \hspace{0.1in}
    \begin{subfigure}{0.44\linewidth}
        \includegraphics[width=\linewidth]{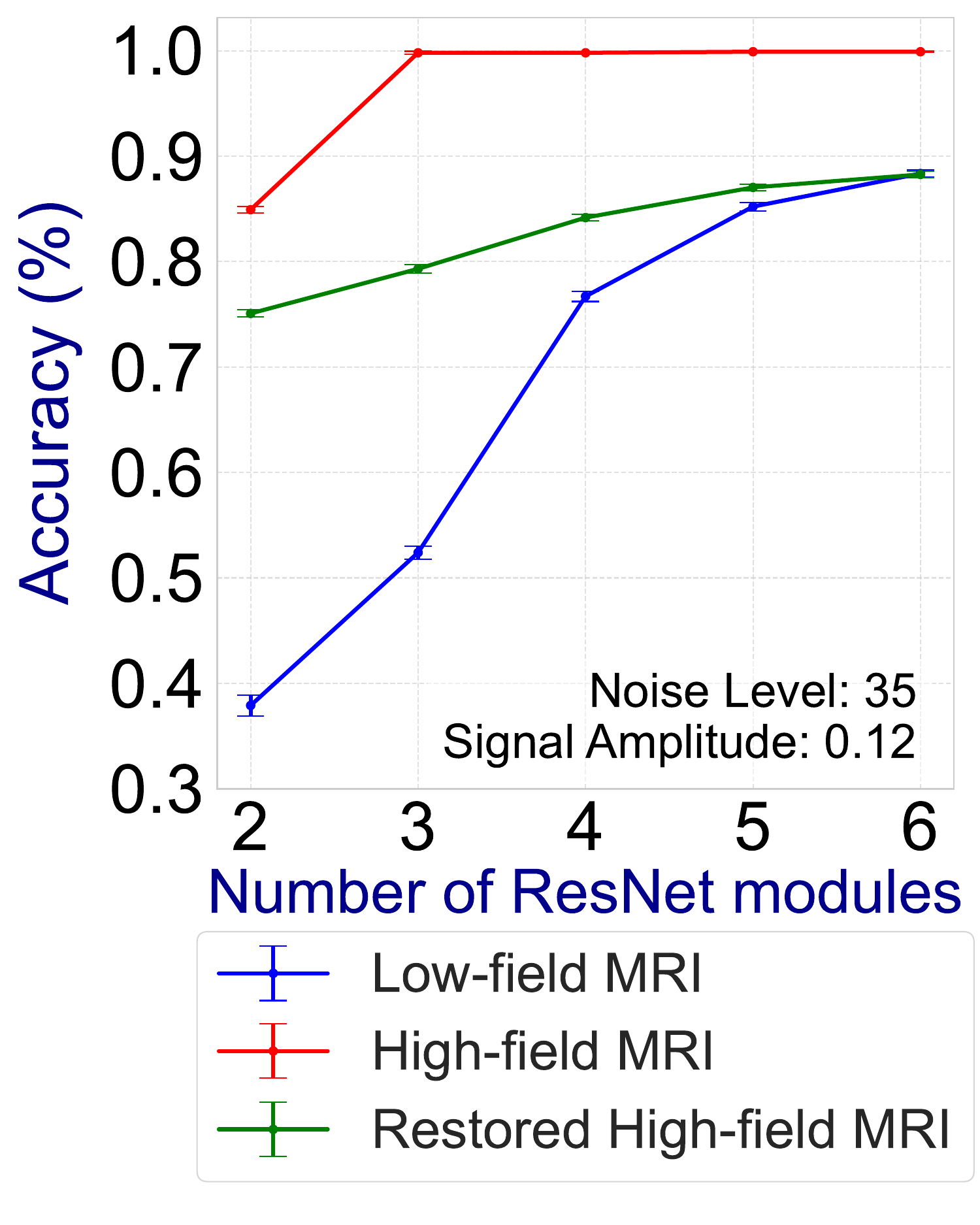}
        \caption{}
    \end{subfigure}
    \caption{Observer performance on the three-class classification task using a balanced dataset as quantified by $\mathcal{V}$-info (a) and classification accuracy (b), shown as functions of observer capacity and background/signal settings. 
    Both metrics exhibit consistent trends of performance improvement with increasing observer capacity. 
    However, classification accuracy saturates at higher capacities and fails to reflect further improvements (first and second rows), while $\mathcal{V}$-info continues to increase.}
    \label{v-info and accuracy}
    \vspace{0.05in}
\end{figure}

\subsubsection{Studies involving three-class  tasks (balanced data)}

The results for the three-class classification tasks with balanced data are reported here.  Observer performance was evaluated under varied signal and background settings (see Appendix~\ref{signal-setting}) and across five levels of observer capacity, using a balanced dataset with an equal number of images from each class. In these tasks, where AUC is not straightforward to compute, classification accuracy was computed and compared to $\mathcal{V}$-info.

Figure~\ref{v-info and accuracy} shows that both  $\mathcal{V}$-info and classification accuracy displayed consistent trends. Specifically,  classification performance increased as NO capacity increased across all background and signal settings. 
The restored high-field images consistently ranked higher than the low-field images across the entire range of observer capacities. 
As NO capacity increased, both $\mathcal{V}$-info and accuracy computed using the low-field images  improved rapidly, eventually becoming comparable to the values computed using the restored high-field images.
However, accuracy quickly saturated, particularly at higher capacities, while $\mathcal{V}$-info continued to increase. This behavior highlights the enhanced sensitivity of $\mathcal{V}$-info in capturing performance differences that are not strongly conveyed by accuracy.

\begin{figure}[!ht]
  \centering
  \vspace{0.05in}
  \begin{subfigure}{0.32\linewidth}
    \includegraphics[width=\linewidth]{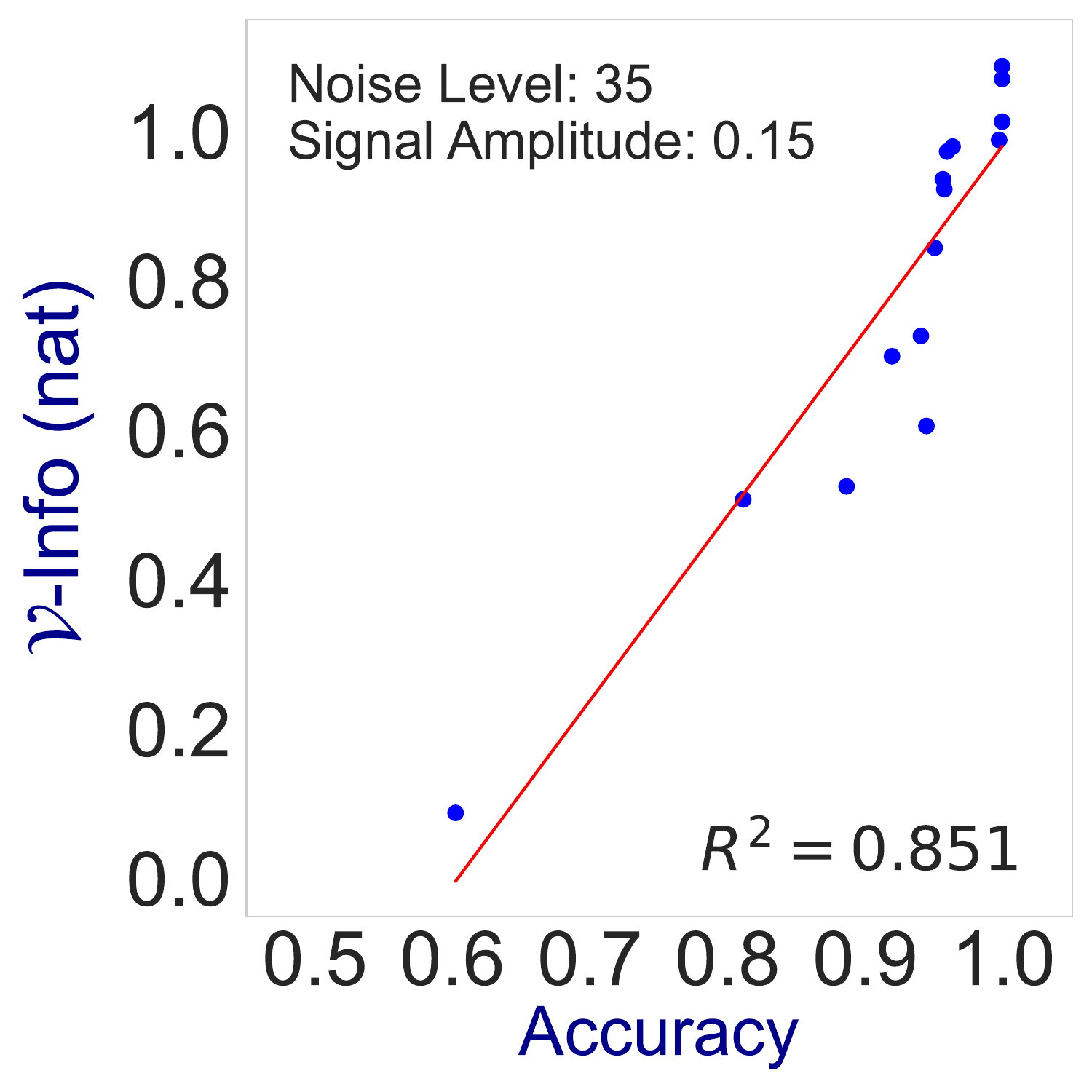}
    \caption{Low-field MRI}
  \end{subfigure}\hfill
  \begin{subfigure}{0.32\linewidth}
    \includegraphics[width=\linewidth]{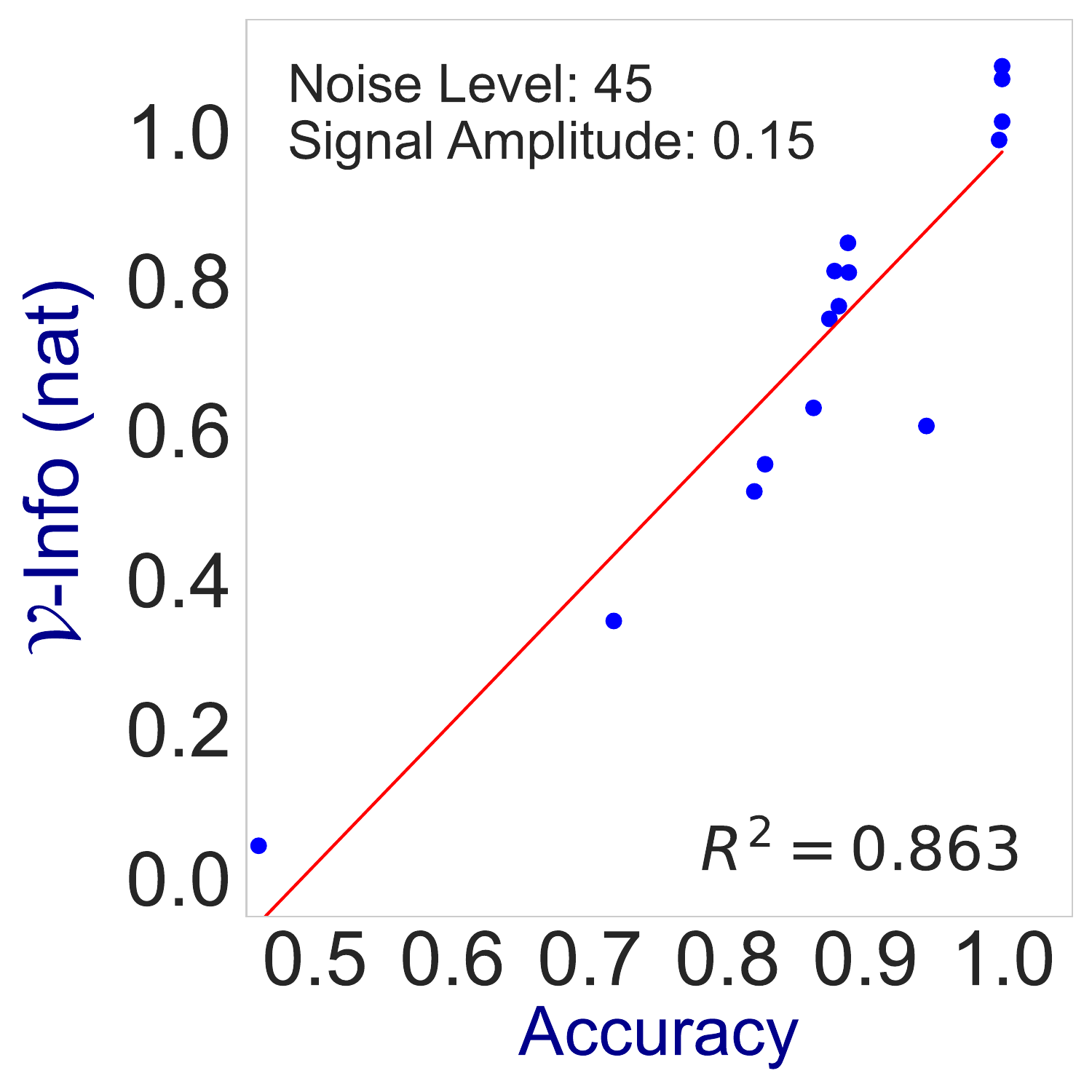}
    \caption{High-field MRI}
  \end{subfigure}\hfill
  \begin{subfigure}{0.32\linewidth}
    \includegraphics[width=\linewidth]{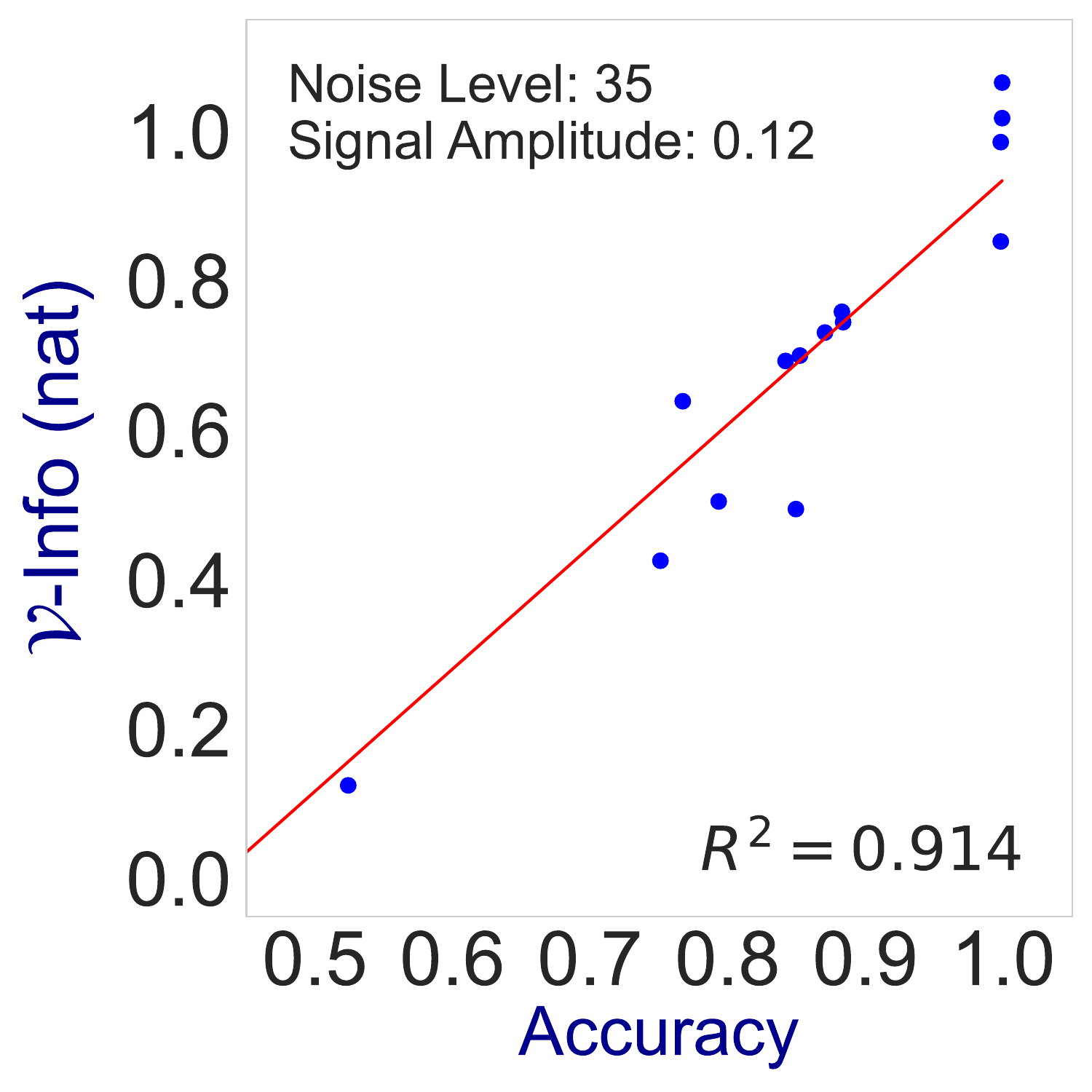}
    \caption{Restored MRI}
  \end{subfigure}
  \caption{Relationship between $\mathcal{V}$-info and Accuracy for ResNet-based observers (three-class classification task). 
  Blue dots denote observer performance at five capacity levels (2–6 ResNet modules) across three imaging conditions (low-field, high-field, restored).  
  The linear trend is slightly weaker than that in Fig.~\ref{vinfo_auc_cnn}, with $R^{2}\!\approx\!0.87$.}
  \label{vinfo_acc_resnet}
\end{figure}

This observation is further supported by a linear regression analysis between accuracy and $\mathcal{V}$-info across all signal and background conditions, as shown in Fig.~\ref{vinfo_acc_resnet}. While a linear correlation existed overall, the determination coefficient $R^2 \approx 0.87$ was notably lower than that observed for the binary detection task. This suggests that accuracy became less reliable in differentiating observer performance as it approaches its upper bound, whereas $\mathcal{V}$-info maintained discriminative power.

\begin{figure}[H]
    \centering
    \begin{subfigure}{0.42\linewidth}
    \includegraphics[width=\linewidth]{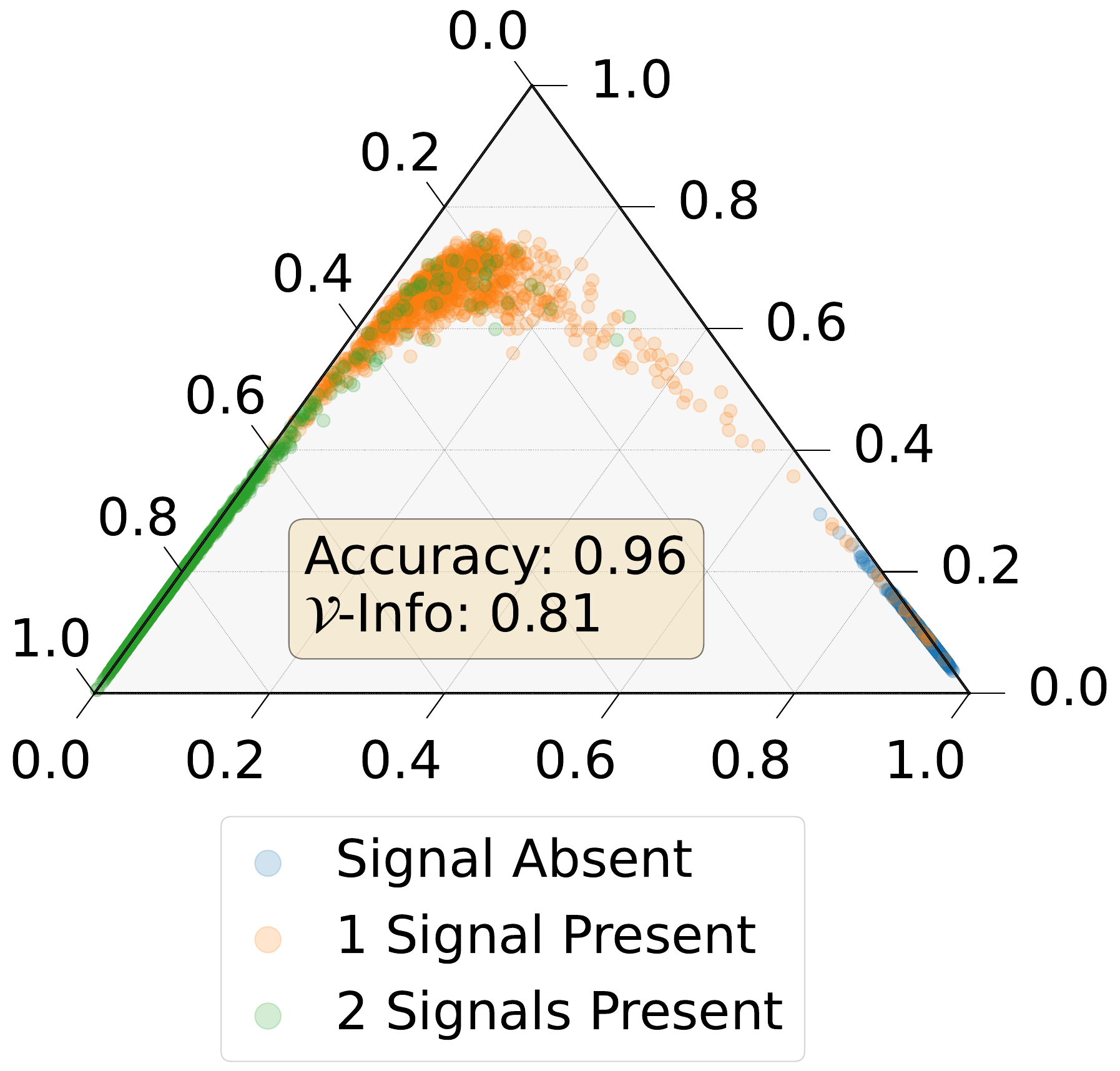}
    \end{subfigure}
     \hspace{0.1in}
    \begin{subfigure}{0.42\linewidth}
        \includegraphics[width=\linewidth]{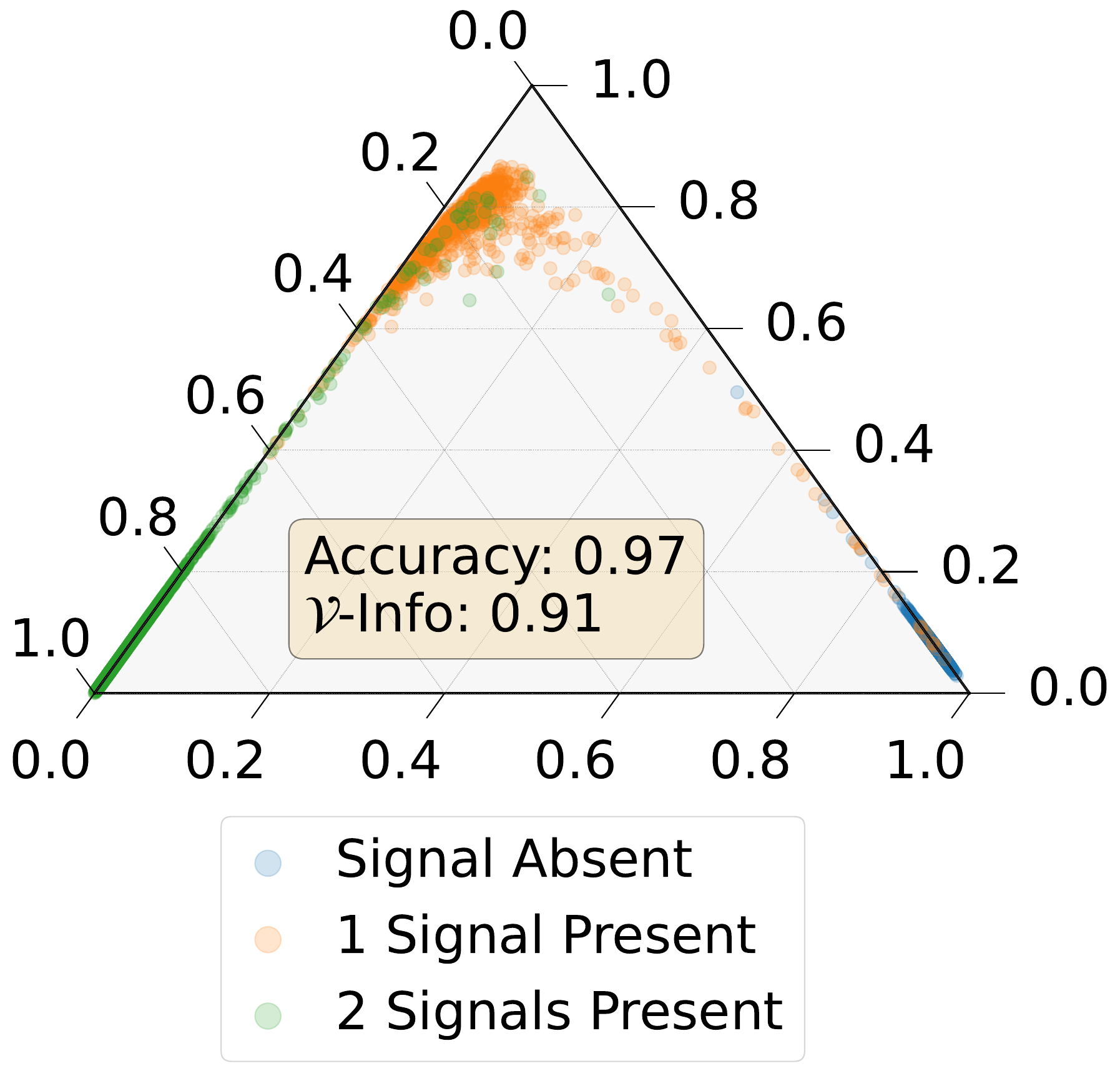}
    \end{subfigure}
    \caption{
Comparison of accuracy and $\mathcal{V}$-info for a three-class classification task using a balanced dataset. Ternary plots show predicted class probabilities, where each point represents a prediction positioned according to its probabilities across the three classes (signal absent, one signal present, and two signals present) and colored by the true label. While accuracy values remain comparable across the two examples, $\mathcal{V}$-info reveals a larger difference, suggesting greater sensitivity to variations in the predicted probability distributions.
}
    \label{tenary plot}
\end{figure}

In addition, Fig.~\ref{tenary plot} provides a stylized example illustrating the sensitivity of $\mathcal{V}$-info over accuracy. Specifically, the task considered was a three-class classification problem, where three predicted probability distributions with different central tendency and dispersion were compared. 
Between two different training checkpoints of the same model,
accuracy increased by only $1\%$ (from 0.96 to 0.97) while $\mathcal{V}$-info increased by $12\%$ (from 0.81 to 0.91). 
This demonstrates that $\mathcal{V}$-info is more sensitive to changes in predictive distribution and can capture performance variations that are overlooked by conventional metrics such as accuracy.

\subsubsection{Studies involving imbalanced data}

A study was performed to investigate the impact of imbalanced training data on task performance as measured by $\mathcal{V}$-info, AUC, and accuracy. 
The image restoration network, CNN-based NOs, and ResNet-based NOs were retrained using the imbalanced training data described in Sec.~\ref {NO-Computation}. Additional details are provided in  Appendices~\ref{Restoration-training} and \ref{NO-setting}.

\begin{figure}[!h]
   \centering
   \begin{subfigure}{0.45\linewidth}
        \includegraphics[width=\linewidth]{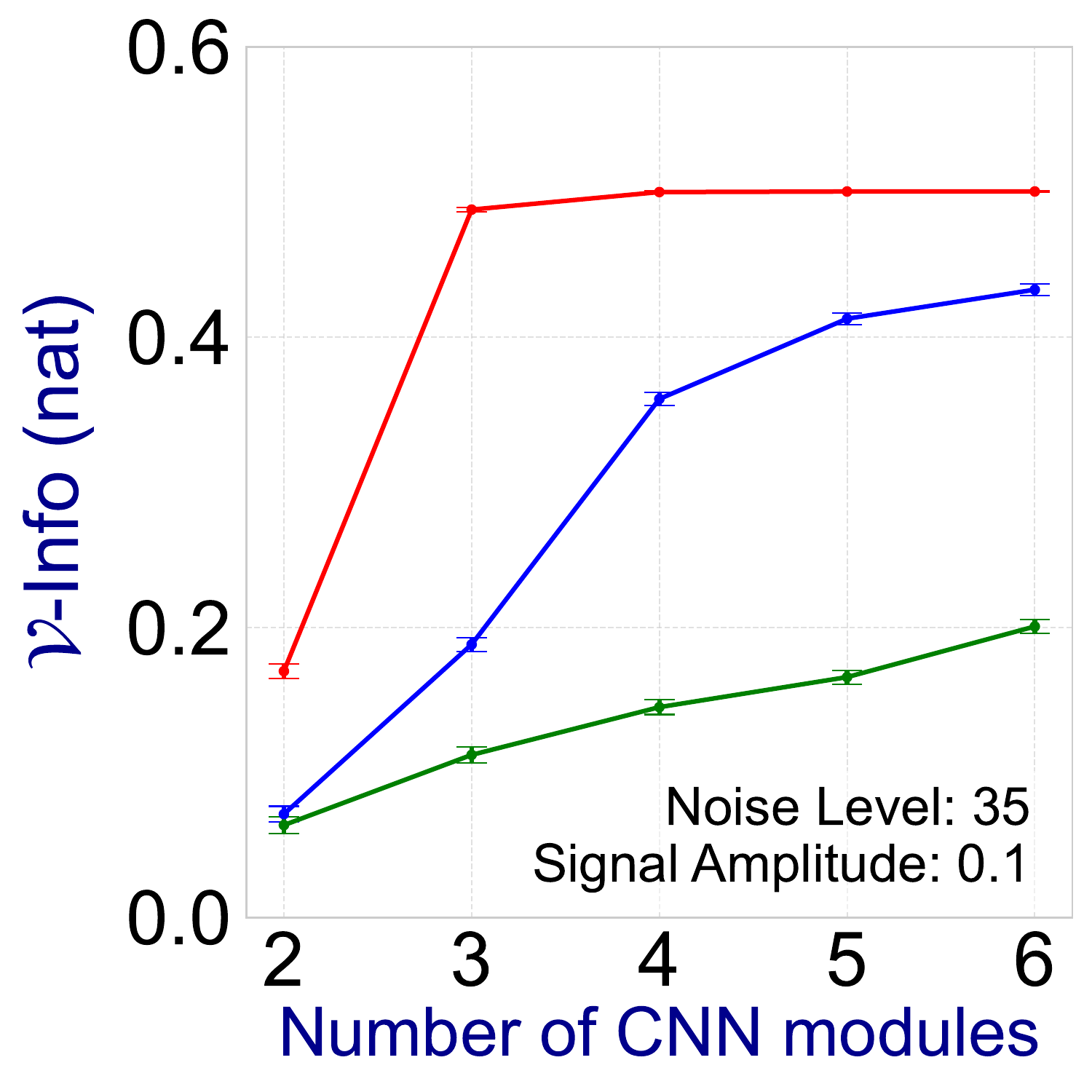}
    \end{subfigure}
        \hspace{0.1in}  
    \begin{subfigure}{0.45\linewidth}
        \includegraphics[width=\linewidth]{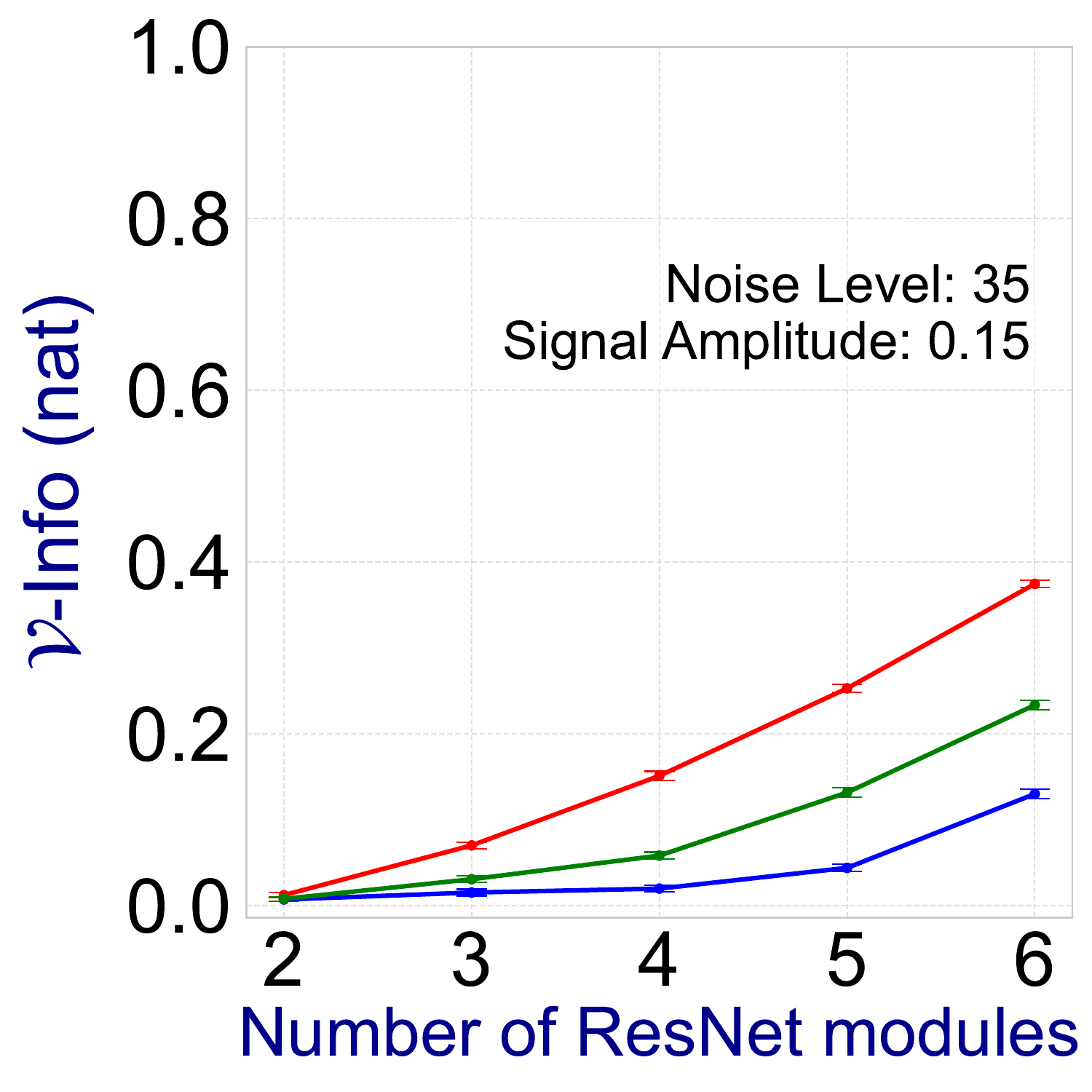}
    \end{subfigure} 
    
    \begin{subfigure}{0.45\linewidth}
        \includegraphics[width=\linewidth]{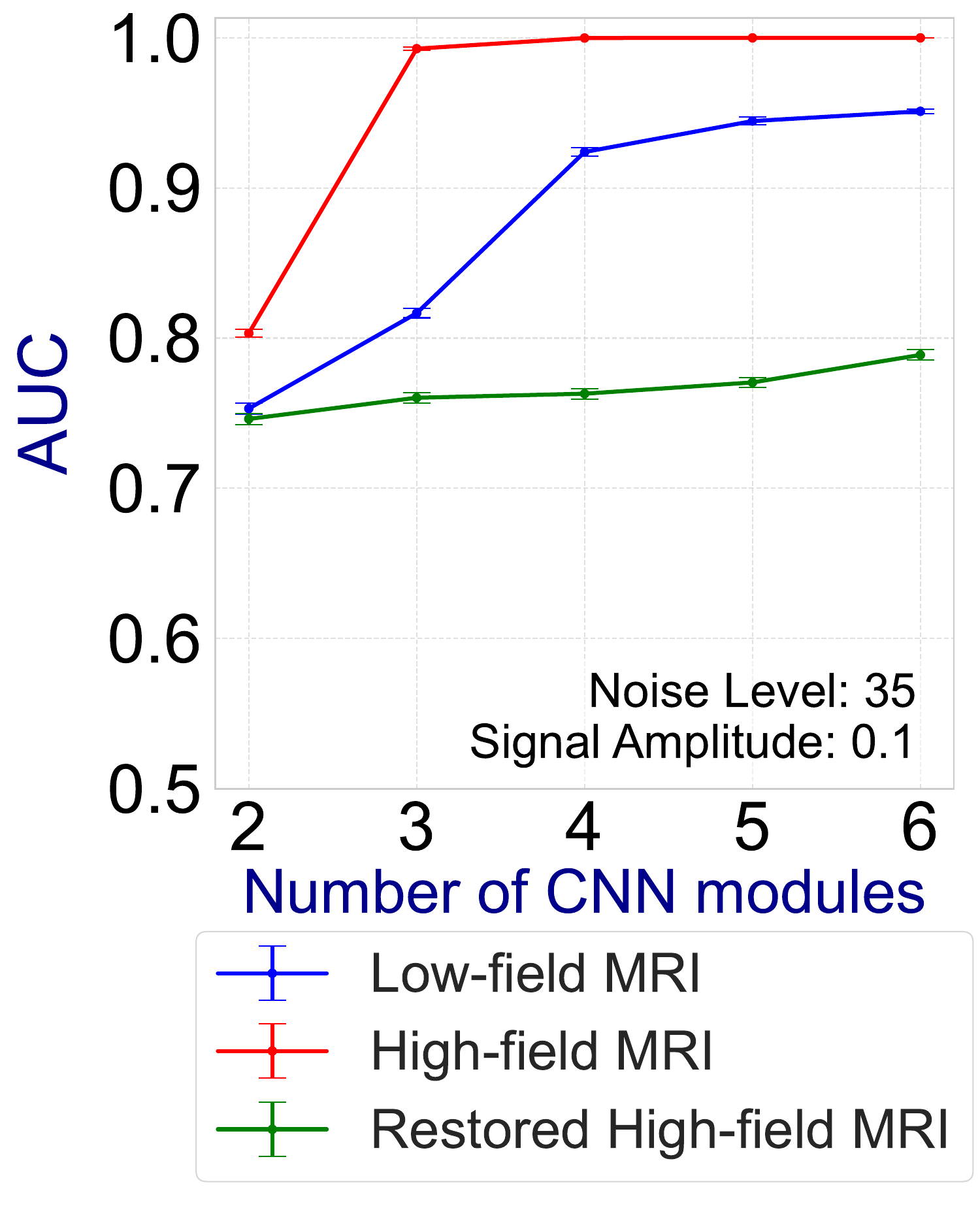}
        \caption{CNN-based Observers}
    \end{subfigure}
       \hspace{0.1in}
    \begin{subfigure}{0.45\linewidth}
        \includegraphics[width=\linewidth]{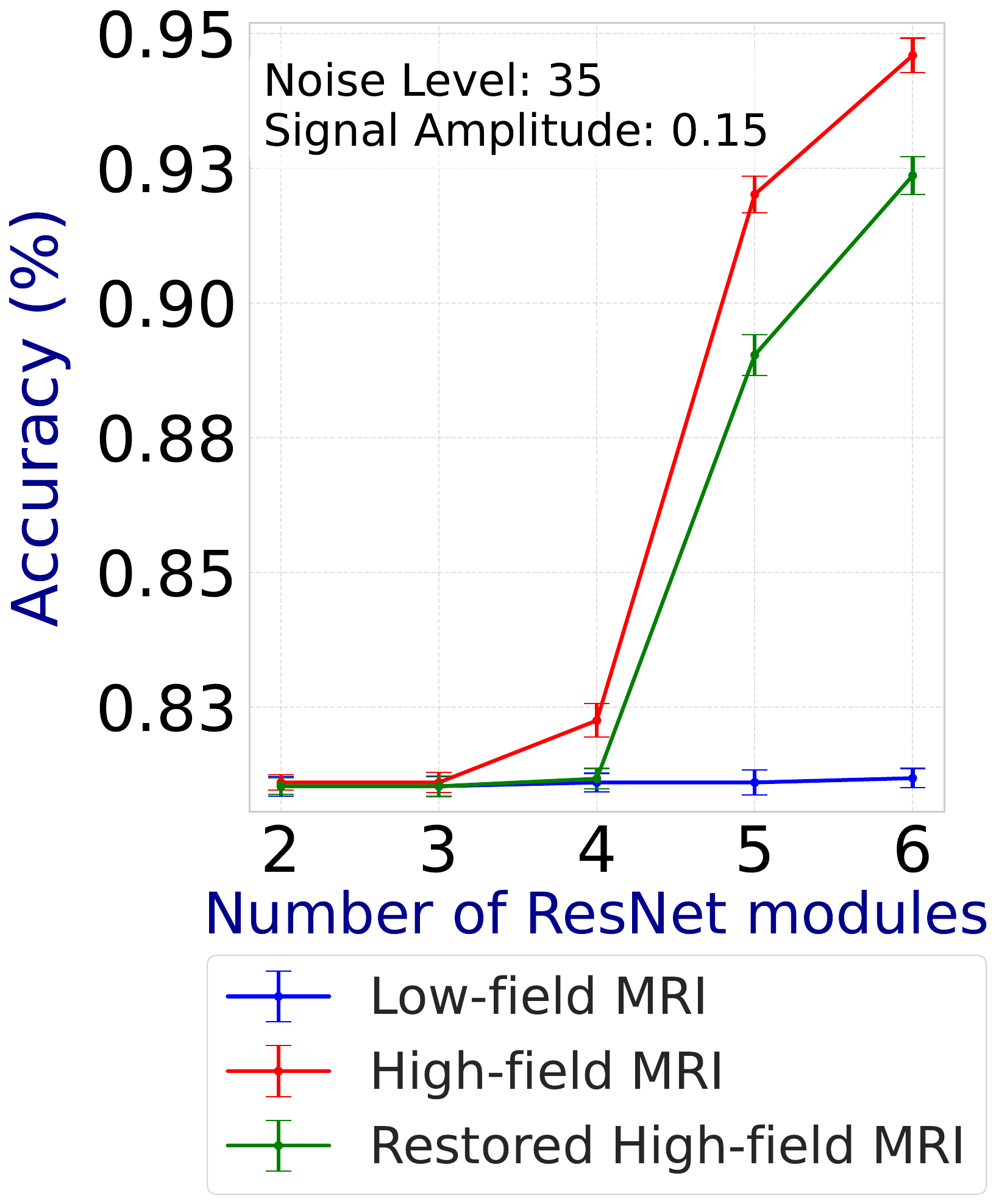}
        \caption{ResNet-based Observers}
    \end{subfigure}
    \caption{
    Effect of imbalanced training data on observer performance: $\mathcal{V}$-info vs. AUC for (a) binary detection  and (b) $\mathcal{V}$-info vs. accuracy for three-class classification. } 
    \label{imbalanced data}
    \vspace{-0.1in}
\end{figure}

As shown in Fig.~\ref{imbalanced data}, for both binary detection and three-class classification tasks, AUC, accuracy, and $\mathcal{V}$-info increased with model capacity across all image types, background, and signal settings. For the binary signal detection task, increasing the number of CNN modules in the CNN-based NO from 5 to 6 improved $\mathcal{V}$-info by $3.9\%$ on the low-field images, while AUC increased by only $1.3\%$. On restored high-field images, the same change led to a $\mathcal{V}$-info gain of $7.0\%$, compared to an AUC gain of $3.6\%$. 
For the three-class classification task, increasing the number of ResNet modules in the ResNet-based NO from 5 to 6 led to a $13.4\%$ increase in $\mathcal{V}$-info on low-field images, while accuracy improved by only $0.4\%$. On restored high-field images, the same module change led to a $19.0\%$ increase in $\mathcal{V}$-info, whereas accuracy increased by $16.6\%$.
These observations made under imbalanced data conditions confirm that $\mathcal{V}$-info remains a robust and sensitive measure of observer performance.
{It is noteworthy that an AUC of 0.5 and $\mathcal{V}$-info of 0 represent the random guess baselines for the performance of observers trained with either balanced or imbalanced data. However, accuracy is sensitive to class distribution, and its baseline under random guessing shifts with imbalance—undermining its reliability as a performance metric.

The relationship between $\mathcal{V}$-info and AUC/accuracy under imbalanced data conditions was further investigated and compared to that observed with balanced data. As shown in Fig.~\ref{relation2}, strong and comparable linear relationships exist between $\mathcal{V}$-info and AUC under balanced and imbalanced data conditions for the case of binary signal detection.
In the three-class signal classification case, the correlation coefficient between accuracy and  $\mathcal{V}$-info decreases slightly under imbalanced data conditions compared to the balanced case. 

\begin{figure}[]
    \centering
    \begin{subfigure}{0.45\linewidth}
        \includegraphics[width=\linewidth]{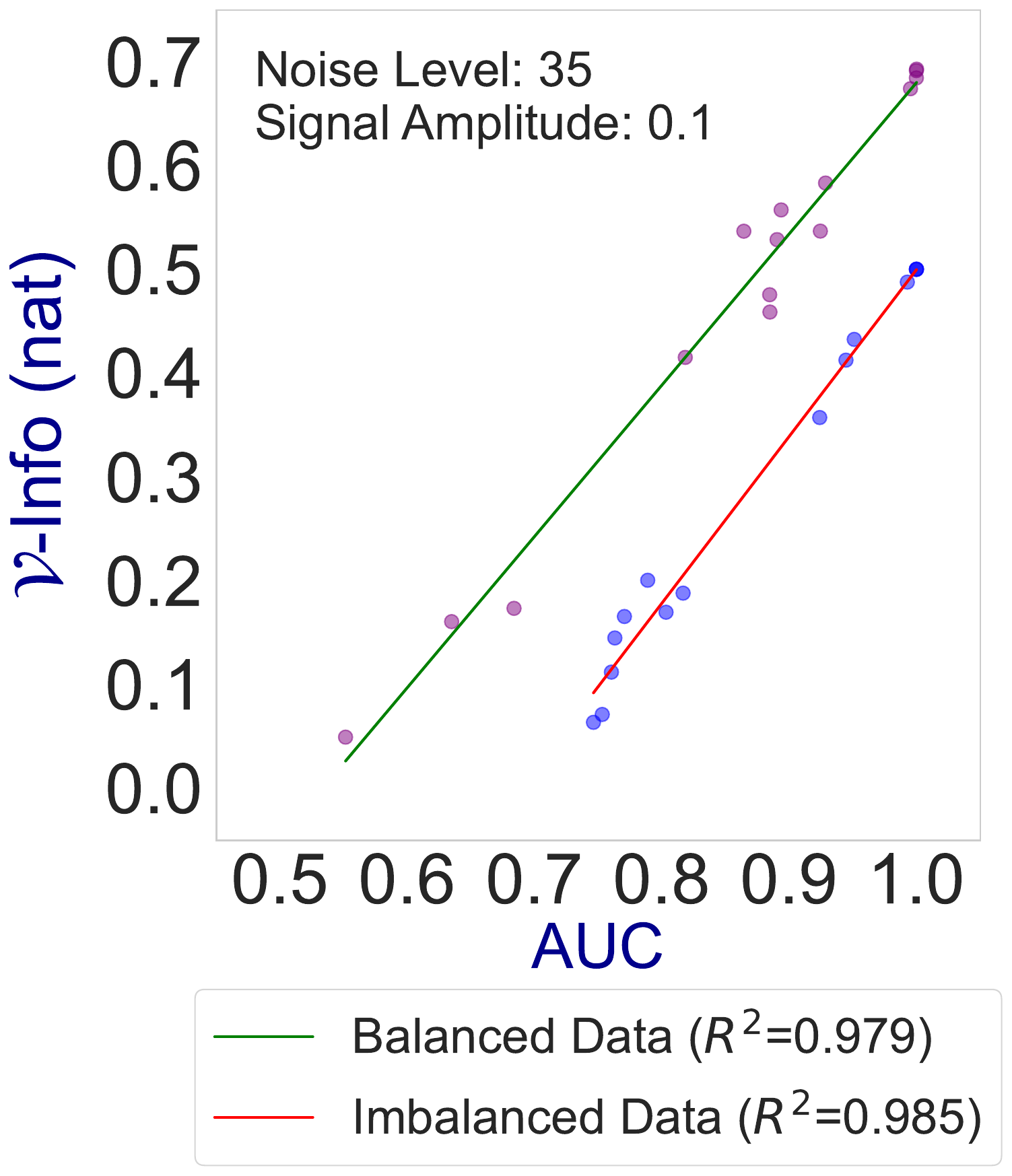}
    \end{subfigure}
     \hspace{0.1in}
    \begin{subfigure}{0.45\linewidth}
        \includegraphics[width=\linewidth]{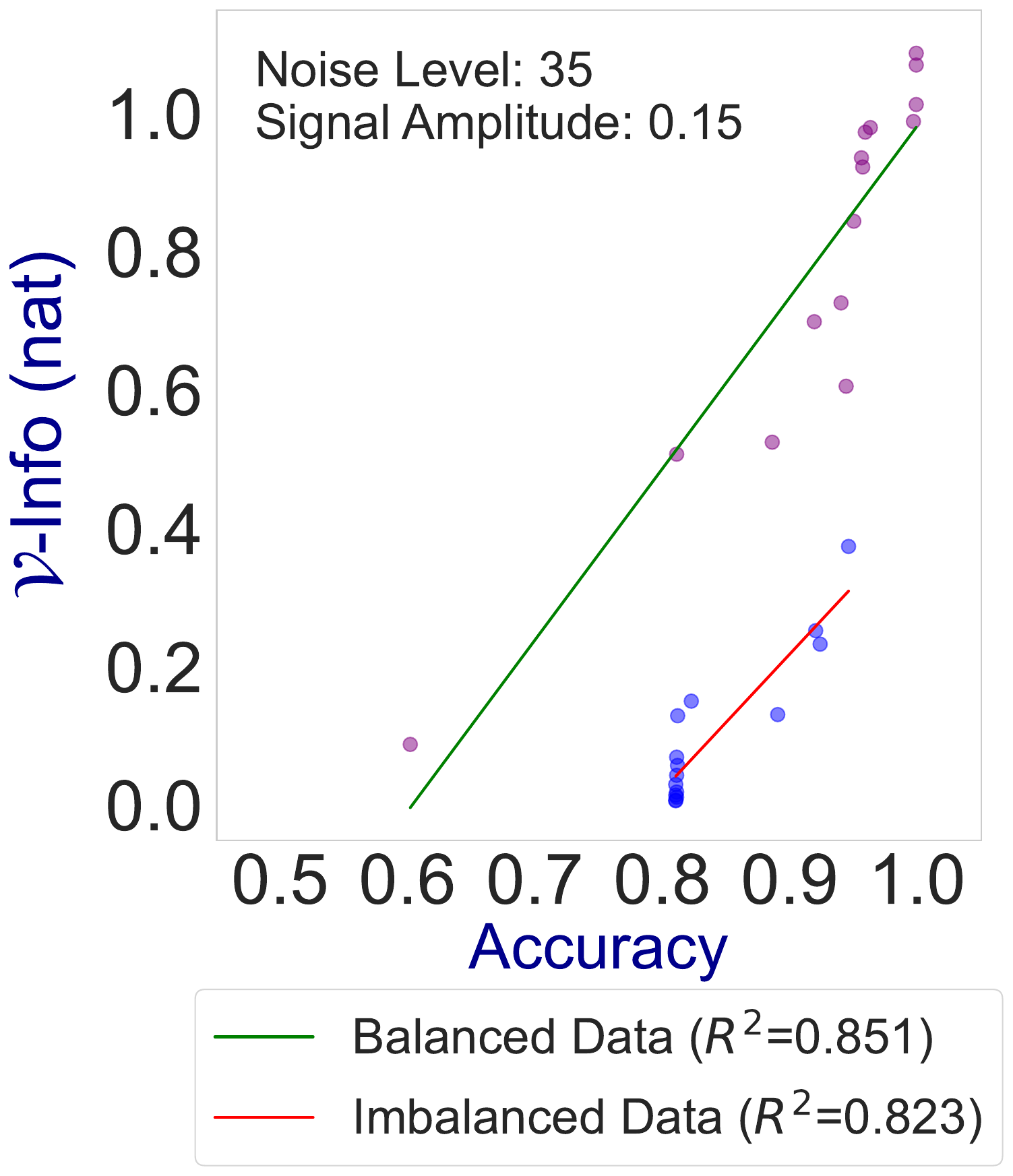}
    \end{subfigure}
    \caption{Relationship between $\mathcal{V}$-info and AUC (left) or accuracy (right) under balanced and imbalanced data conditions for binary detection and three-class classification tasks. Linear regression lines are shown for the balanced data condition (green) and imbalanced data condition (red), with corresponding linear regression coefficient $R^2$ values indicated in the legend. Purple and blue dots represent the observer performance achieved for five levels of observer capacity (2-5 CNN or ResNet modules), and evaluated on low-field, high-field, and restored MRI images for balanced and imbalanced data conditions, respectively. $\mathcal{V}$-info exhibits a strong linear relationship with AUC, while its correlation with accuracy is weaker, particularly under imbalanced conditions.}
    \label{relation2}
       \vspace{-0.1in}
\end{figure}

\vspace{-0.1in}
\section{Discussion and Summary}
\label{discussion}

{ 
Task-based image quality assessment in medical imaging has long relied on
signal detection theory–based observer models, including both ideal and
sub-ideal numerical observers, to relate image quality to task performance.
Well-established numerical observers, including the Hotelling observer and related
variants, explicitly incorporate observer limitations and have proven highly
valuable for predicting detection and discrimination performance under controlled
task formulations. These approaches provide deep insight into how image
statistics, noise correlations, and observer structure influence decision
performance for specific tasks.

$\mathcal{V}$-information is not intended to replace such signal detection theory-based
approaches, but rather to complement them by addressing a different level of
abstraction in task-based image quality assessment. Detection theory-based observers
are typically evaluated through decision-level performance metrics tied to
specific task formulations and rules that specify how different types of prediction errors are
penalized when converting task predictions into decision performance. Alternatively, $\mathcal{V}$-info
quantifies the amount of task-relevant information that is accessible to a given
observer class through its estimated posterior distribution, independent of any
particular decision rule or operating point. This distinction enables
$\mathcal{V}$-info to serve as a unifying, observer-aware measure of image
utility that applies naturally across a broad range of task types, including
multi-class classification tasks. 

By focusing on the information accessible to a specified class of observers rather
than on specific decision outcomes, $\mathcal{V}$-info provides a principled basis
for comparing imaging systems and processing pipelines in settings where observer
implementations and decision strategies may vary within that class, or where
downstream decision rules  have not yet been fully specified.
This perspective is particularly valuable in contemporary medical imaging
workflows, in which observer implementations and downstream uses of image data
often differ across applications and evolve over time. As a result, $\mathcal{V}$-info
supports stable and interpretable system-level comparisons that complement
traditional decision-based metrics by characterizing image utility in a manner that
is less sensitive to specific deployment choices.

In the numerical experiments, $\mathcal{V}$-info was
observed to be empirically correlated with conventional decision-based metrics
such as AUC or accuracy. In particular, increasing the capacity of the classifier
model generally led to concurrent increases in both $\mathcal{V}$-info
and AUC or accuracy. Such behavior is expected in regimes where (i) the observer
class is relatively simple or capacity-limited, and (ii) task difficulty is
moderate and performance is not yet saturated. Under these conditions,
improvements in task performance tend to manifest simultaneously as increases in
both discriminative ability and the amount of task-relevant information
accessible to the observer.
Importantly, such correlation reflects theoretical consistency rather than
redundancy.  While these measures may track similar trends in certain regimes,
they quantify fundamentally different aspects of task-based performance and can
diverge outside of these conditions.
This parallels the well-known behavior between ideal-observer detectability and mutual information.
Moreover, as demonstrated by the results in Section~\ref{results} and Appendices \ref{additional-experiment} and \ref{sensitivity-vinfo}, $\mathcal{V}$-info
remains informative in regimes where decision-based metrics may saturate.

Because $\mathcal{V}$-info is estimated from finite datasets, the resulting
values are subject to statistical uncertainty. Xu et al.\ \cite{xu2020theory}
derived finite-sample error bounds for $\mathcal{V}$-information estimates, which
tighten as the dataset size increases, as expected. Future work should therefore
investigate the behavior of $\mathcal{V}$-info in data-limited regimes and
develop practical methods for associating confidence intervals or uncertainty
estimates with reported $\mathcal{V}$-info values. In addition, a
systematic analysis of computational trade-offs across different observer classes
and data scales remains an important direction for future research, particularly
for large three-dimensional or multi-modal datasets.

Beyond classification tasks, the $\mathcal{V}$-info framework naturally
extends to a broader class of inference problems. Although the task variable is
instantiated as a categorical label in the experiments presented here, the
framework itself does not rely on hard labels (i.e., deterministic task annotations that assign a single outcome rather than a
distribution over possible outcomes) or discrete decision outcomes.
Instead, it accommodates general task variables defined through conditional
distributions. This includes estimation tasks, detection–estimation hybrid tasks,
and structured prediction problems.
Exploring these extensions represents an
important direction for future work.

 {The explicit purpose of the predictive function family $\mathcal{V}$ is precisely to represent
observer-specific behavior, including anthropomorphic and clinically relevant observers. While the present study focuses on establishing a general and principled framework, future extensions could explicitly model clinically relevant observers by defining observer-specific choices of $\mathcal{V}$ that more closely model
human reader performance.
Such observer-specific models could be informed by psychophysical studies, eye-tracking data, or data-driven approaches trained on expert annotations.}

In summary, $\mathcal{V}$-information represents a theoretically grounded
task-based image quality measure that quantifies how much information about a
specified task is accessible to a given class of observers. By explicitly
incorporating observer limitations through the choice of the predictive class
$\mathcal{V}$, it provides an observer-aware assessment of image utility based on
the observer’s estimated posterior distribution of the task variable. As a
result, $\mathcal{V}$-info  reflects not only whether task-relevant
features are present in the image, but also how much uncertainty about the task
variable remains after the image has been observed, and thus how reliably the
image supports task-related conclusions.
}

\vspace{-0.1in}
\section{Acknowledgments}
This work was supported in part by NIH Awards P41EB031772 (sub-project 6366), R01EB034249, R01CA233873, R01CA287778, and R56DE033344.
\vspace{-0.1in}

\bibliographystyle{IEEEtran} 
\bibliography{report} 

\vspace{-0.1in}
\appendix
\vspace{-0.05in}

\subsection{MRI Restoration Architecture Details}
\label{restoration-details}
As depicted in Fig.~\ref{unet}, the U-Net architecture comprises an encoder and a decoder, containing four downsampling and four upsampling convolutional blocks, respectively. The encoder channels progressively increase as \(1, 64, 128, 256, 512\), while the channels in decoder progressively decrease as \(512, 256, 128, 64, 1\). Before the first downsampling step in the encoder, a Conv+ReLU layer is applied, in which 64 convolutional filters of size \(3 \times 3 \times 1\) are employed to generate 64 feature maps. Similarly, in the decoder, after the final upsampling step, a single convolutional filter of size \(3 \times 3 \times 64\) is applied to produce the final estimated high-field MRI with dimensions \(288 \times 320\). 

\vspace{-0.15in}
\subsection{Image Restoration Network Training Details}
\label{Restoration-training}

Each group of low-field and high-field MRIs contains 11,130 MRI images. These images were further divided into training, validation, and testing datasets in an 8:1:1 ratio. 
All restoration networks were trained using the Adam optimizer with a learning rate of 0.00005 for 100 epochs with an early stopping rule. Mini-batches of 32 images were utilized at each training iteration. 
The loss function is defined in \eqref{mse}.
The network model that achieved the best performance on the validation dataset was selected for further evaluation. All networks were implemented and trained using the PyTorch framework on four NVIDIA A40 GPUs.  
To assess the robustness of the restoration process under data imbalance conditions, 8,906 signal-absent images and 1,781 signal-present images were employed for restoration model training for the binary task.
For the three-class classification task, 8,906 signal-absent images, 1,781 one-signal-present, and 1,781 two-signals-present images were employed for model training. The validation and testing datasets were kept the same as in the balanced setting to ensure fair evaluation.

\vspace{-0.1in}
\subsection{Signal and Background Image Hyperparameters Settings}
\label{signal-setting}

The background noise level was defined by the standard deviation ($\sigma$) of Gaussian noise.
For the binary signal detection task, three background noise and signal settings, $(\sigma=35, \text{amplitude}=0.1)$, $(\sigma=45, \text{amplitude}=0.1)$, and $(\sigma=35, \text{amplitude}=0.08)$ are defined. 
For three-class classification tasks, the three background noise and signal settings are $(\sigma=35, \text{amplitude}=0.15)$, $(\sigma=45, \text{amplitude}=0.15)$, and $(\sigma=35, \text{amplitude}=0.12)$, respectively.

\vspace{-0.1in}
\subsection{Numerical Observer Architecture and Training Details}
\label{NO-setting}

\vspace{-0.05in}
\subsubsection{CNN observer}
\label{CNN-IO}
CNN-based observers were used to assess the impact of MRI restoration on binary signal detection. Each observer consisted of stacked convolutional modules with Instance Normalization, ReLU activations, and average pooling for downsampling, with channels progressively increasing to enable multiscale feature extraction. The final feature map was flattened and passed through a fully connected layer to produce binary classification logits. Inputs included restored high-field, high-field, or low-field MRI images of size $288 \times 320$. $\mathcal{V}$-info was computed after optimizing the observer using \eqref{opt-vinfo}, while AUC was obtained by training with cross-entropy loss and evaluating predictions on the test set. Models were selected based on validation performance.

\subsubsection{ResNet observer}
ResNet-based observers were employed to evaluate the impact of MRI restoration on three-class classification tasks. The ResNet-based observer architecture begins with an initial convolutional layer, followed by two groups of residual modules. Each group contains between one and six residual modules, depending on the model configuration. In the first group, the network employs residual blocks with 32 channels, while in the second group, the channel dimensions increase to 64. The spatial resolution is reduced by half at the beginning of the second group using a stride of 2, enabling the extraction of features at multiple scales.

The input image is first processed by a convolutional layer with 32 filters of size 3×3×1, followed by Batch Normalization and a ReLU activation function. Within each residual block, there are two convolutional layers with Batch Normalization and ReLU activations, along with a shortcut connection. The fully connected layer at the end of the network maps the extracted feature vector to the classification output. The output is a three-label probability vector corresponding to the signal-absent, one-signal-present, and two-signal-present. The network input can be an estimated high-field MRI, a high-field MRI, or a low-field MRI image. 
$\mathcal{V}$-info is computed directly after optimizing the observer using \eqref{opt-vinfo}. 
In contrast, AUC is computed by first training the observer on the training set using cross-entropy loss, followed by evaluating the model’s predictions on the test set.
The observers used for evaluation are selected based on their best performance on a separate validation set.

\subsection{Details of AUC/Accuracy and $\mathcal{V}$-info Plot}
In Fig.~\ref{binary plot}, the left panel illustrates the probability distributions obtained in the early stage of training (signal strength $=0.1$, background noise $=35$, on restored high-field image), while the right panel corresponds to the mid-training stage. Similarly, in Fig.~\ref{tenary plot}, the left panel shows the ternary probability distributions in the early stage of training (signal strength $=0.15$, background noise $=35$, on restored high-field image), and the right panel depicts the corresponding mid-training distributions. All AUC and accuracy values were obtained on the testing set, $\mathcal{V}$-info was obtained on the training set.

\subsection{Numerical Studies using clinical imaging datasets.}
\label{additional-experiment}
{ 
We conducted additional supplementary experiments on real-world medical
imaging datasets over two modalities and three classification tasks.
Specifically, we employed a chest radiography dataset CheXpert \cite{irvin2019chexpert} and an optical coherence tomography dataset, Kermany OCT\cite{kermany2018identifying}.

\emph{Task 1} (Normal vs.\ Pneumonia, CheXpert) is a binary thoracic disease classification task distinguishing normal chest radiographs from the Pneumonia class using CheXpert dataset. The training set contained 15,288 normal images and 4,195 Pneumonia images. The validation set contained 853 normal images and 229 Pneumonia images, while the test set contained 832 normal images and 251 Pneumonia images.

\emph{Task 2} (Normal vs.\ lung-lesion, CheXpert) is a binary classification task distinguishing normal chest radiographs from those labeled with a lesion using CheXpert dataset. The training set included 15,278 normal images and 6,332 lesion images. The validation set contained 845 normal images and 356 lesion images, and the test set contained 850 normal images and 351 lesion images.

\emph{Task 3} (4-class retinal disease, OCT) is a four-class retinal disease classification task using the Kermany OCT dataset to differentiate normal, diabetic Macular Edema (DME), Choroidal neovascularization (CNV), and drusen. The training set contains 37,000 CNV, 11,000 DME, 8,600 drusen, and 26,300 normal images. Both the validation and test sets include 242 images per class.

All experiments employed a consistent suite of observer models, including three convolutional neural network (CNN) architectures with progressively increasing depth, as well as EfficientNet-B0 and EfficientNet-B5 \cite{tan2019efficientnet}. The CNN variants follow the architectural design described in Appendix~\ref{CNN-IO} with {CNN2}, {CNN3}, and {CNN4} corresponding to networks composed of 2, 3, and 4 convolutional modules, respectively. 
Together, these five models are ordered by increasing architectural depth and number of parameters - CNN2, CNN3, CNN4, EfficientNet-B0, and EfficientNet-B5.

Representative images from these three clinical tasks are shown in
Fig.~\ref{clinical-task}. The quantitative results comparing traditional performance measures (AUC and F1-score) with $\mathcal{V}$-info are presented in
Fig.~\ref{clinical-task-results}. Across all three tasks, $\mathcal{V}$-info consistently increased with observer capacity from CNN2 to CNN4 and Eff0 to Eff5. In contrast, traditional performance measures exhibited saturation in certain cases. For example, the AUC plateaued in the normal vs. lung-lesion classification task (Task 2). For the four-class retinal disease classification (Task 3), the average F1 score computed across all classes saturates beyond the CNN3 architecture. 
This indicates that $\mathcal{V}$-info captures differences in task-relevant information even when standard performance metrics no longer improve. Taken together, these results obtained on real clinical
datasets exhibit trends consistent with those observed in Section~\ref{results} on the MRI simulation study, further illustrating the practical behavior of $\mathcal{V}$-information
across modalities and task settings. These results highlight the general applicability of the $\mathcal{V}$-info framework by construction.
}

\textbf{Statistical significance of sensitivity comparison:}
 {For the OCT retinal four-class classification task, each network depth (2, 3, and 4) was trained using 20 independent random initializations, and sensitivity to model capacity was computed independently for each seed by estimating the slope of the performance metric as a function of network depth, yielding paired sensitivity values for $\mathcal{V}$-information and the F1 score under identical stochastic training conditions. The mean sensitivity for $\mathcal{V}$-information was $0.06 \pm 0.007$, while the corresponding value for the F1 score was $0.01 \pm 0.002$, indicating an approximately $3.7\times$ larger sensitivity for $\mathcal{V}$-information. To determine whether this difference is statistically significant, a paired $t$-test was performed across random seeds, and the null hypothesis of equal mean sensitivity was rejected with $p = 2.03 \times 10^{-10}$, demonstrating that the observed increase in sensitivity is statistically robust and not attributable to stochastic variability in training. These results formally confirm that $\mathcal{V}$-information responds more strongly to changes in model capacity than conventional task performance metrics and therefore provides a significantly more sensitive measure for detecting performance differences in this setting.}

\textbf{Computational cost analysis:}
 {The definition of $\mathcal{V}$-information involves two terms: the label
entropy $H(Y)$ and the conditional $\mathcal{V}$-entropy
$H_{\mathcal{V}}[Y\mid X]$. The label entropy $H(Y)$ is estimated directly
from the empirical label distribution and incurs negligible computational
cost. Consequently, the dominant computation arises from estimating
$H_{\mathcal{V}}[Y\mid X]$ in Eq.~(6), which requires solving the same
optimization problem used to train the observer model. As a result, the
computational complexity of estimating $\mathcal{V}$-information is of the
same order as training a standard classification network used for
conventional task-performance evaluation. To quantify this cost, we measured
the wall-clock training time for each task; the results are reported in
Table~\ref{tab:vinfo_runtime}. All models were trained on a Titan X Pascal
GPU. Because $H_{\mathcal{V}}[Y\mid X]$ is obtained through the same training
process used for the task network, evaluating $\mathcal{V}$-information does
not require an additional training stage beyond that already needed to train
the observer model. Therefore, computing $\mathcal{V}$-information introduces
negligible additional computational overhead relative to standard
task-network training.}

\begin{table}[H]
\centering
\caption{Per-epoch runtime for conditional $\mathcal{V}$-entropy estimation. Times are reported as mean per epoch (in minutes) across runs.}
\label{tab:vinfo_runtime}
\setlength{\tabcolsep}{6pt}
\begin{tabular}{lccccc}
\hline
\textbf{Task} & \textbf{CNN2} & \textbf{CNN3} & \textbf{CNN4} & \textbf{Eff0} & \textbf{Eff5} \\
\hline
Normal vs.\ Pneumonia      & 3 & 3.5 & 4 & 7 & 14 \\
Normal vs.\ Lung lesion    & 3.5 & 4 & 4.5 & 7.5 & 14.1 \\
Retinal OCT (4-class)      & 5.2 & 7 & 9 & 11.4 & 20 \\
\hline
\end{tabular}
\end{table}

\begin{figure}[h]
    \centering
\includegraphics[width=0.95\linewidth]{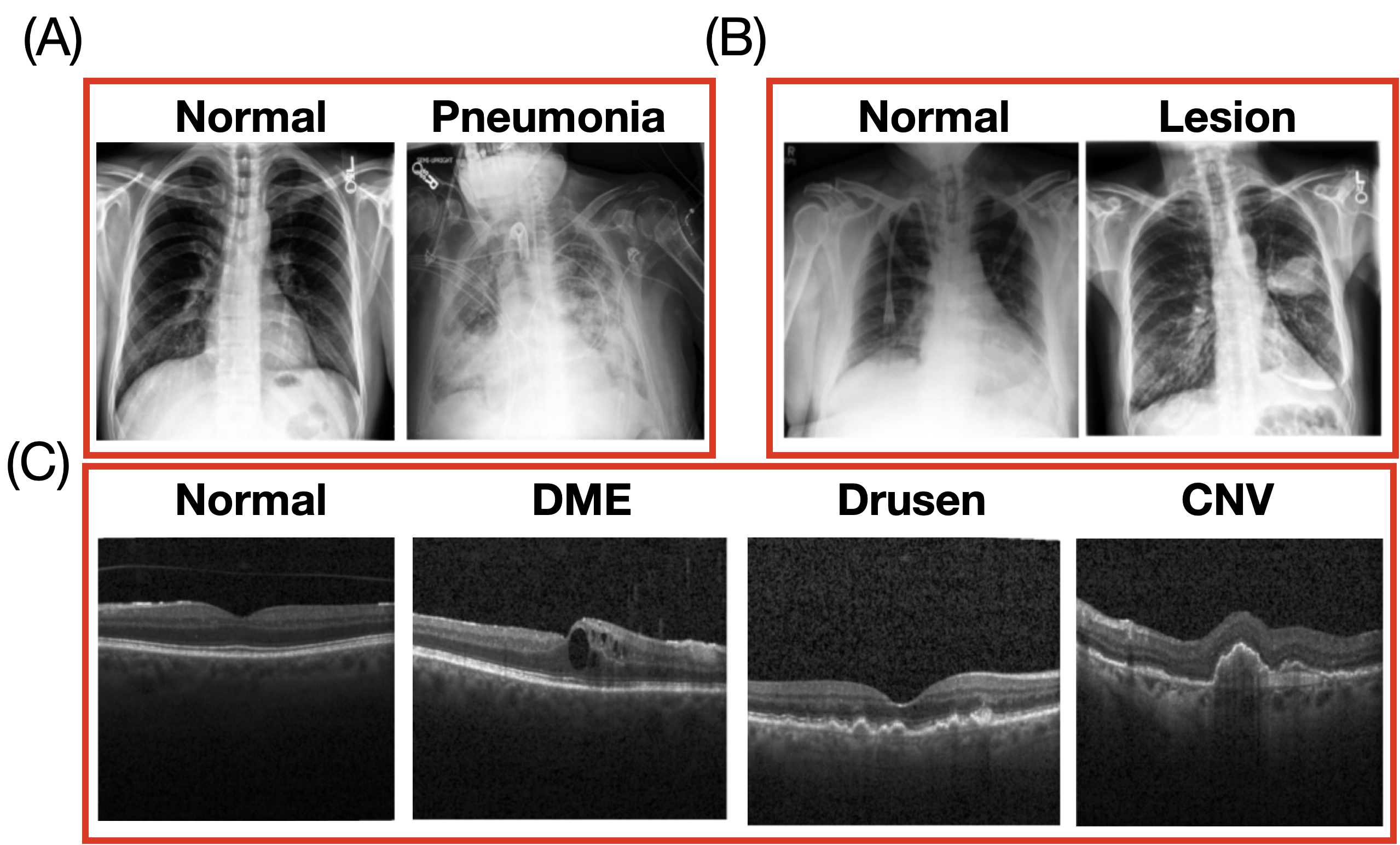}
    \caption{{  Representative examples images from the three additional clinical tasks evaluated in our experiments: (A) CheXpert normal vs. Pneumonia;  (B) CheXpert normal vs. lung-lesion; and (C) Kermany OCT four-class retinal disease classification. 
    }}
    \label{clinical-task}
\end{figure}

\begin{figure}[h]
    \centering
\includegraphics[width=0.95\linewidth]{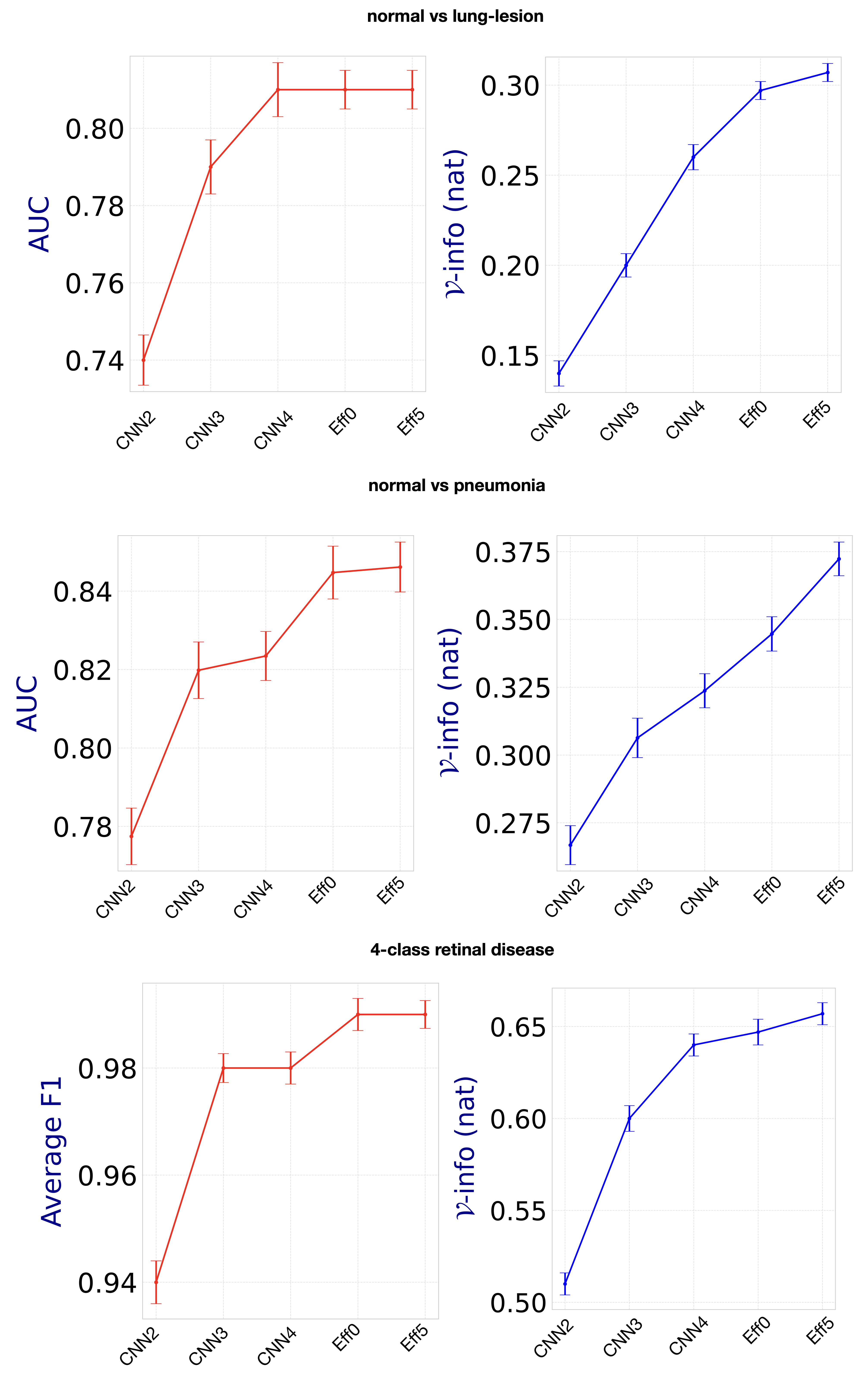}
    \caption{{ 
    Observer performance on three different tasks as quantified by $\mathcal{V}$-info, AUC and F1-score. CheXpert normal vs. Pneumonia, CheXpert normal vs. lung-lesion, and OCT four-class retinal disease classification. 
    Across all three tasks, $\mathcal{V}$-info follow {s} a similar trend like the other traditional measures. Traditional performance measures exhibited saturation in certain cases. For example, the AUC plateaued in the normal vs. lung-lesion classification task (Task 2) and the average F1 score saturates for the four-class retinal disease classification (Task 3). But $\mathcal{V}$-info did not saturate and increased with increasing task network capacity (e.g. - CNN3 to CNN4) 
    }}
    \label{clinical-task-results}
\end{figure}

\subsection{Saturation Behavior Analysis of AUC and $\mathcal{V}$-info}
\label{sensitivity-vinfo}


{ 
\subsubsection{AUC saturation}
As described in (\ref{auc-eq}), for a binary task $\mathbf{Y}\in\{0,1\}$, $\mathrm{AUC}=\Pr\big(s(\mathbf{X}^+)>(\mathbf{X}^-)\big)$. 
Here $s(\cdot)$ denotes an observer’s scalar response or score, and $\mathbf{X}^+$ and $\mathbf{X}^-$ denote image data corresponding to positive and negative task outcomes, respectively. AUC admits the probabilistic interpretation,
and depends only on the ranking induced by $s(\cdot)$. 
Any improvement that does not increase this probability $\Pr\big(\cdot)$ cannot increase AUC.
For any strictly increasing function $g$, the composite function $g\circ s$ satisfies,
\begin{equation}
s(\mathbf{X}^+) > s(\mathbf{X}^-)\iff g\circ s(\mathbf{X}^+) > g \circ s(\mathbf{X}^-).
\end{equation}
Then, 
\begin{equation}
\mathrm{AUC}(s)=\mathrm{AUC}(g\circ s), \\
\end{equation}
Any changes to the numerical values of the predictions that preserve ordering
cannot change AUC. This immediately explains loss of sensitivity: once an
observer is rich enough to produce a response whose ordering is (nearly)
Bayes-optimal, further improvements that refine the posterior probabilities but
do not change the ordering leave AUC unchanged.

\subsubsection{Sensitivity of $\mathcal{V}$-info}

Consider the definition of $\mathcal{V}$-info in (\ref{ouinfo}), where
the involved conditional $\mathcal{V}$-entropy is defined in (\ref{eq:conditional-v-entropy}), in which
$v[x](y)$ represents the observer’s probabilistic
prediction of the task variable given the image data that
approximates the true posterior distribution $p(y \vert x)$.

For a given predictor $v[x](y)$,
the expected negative log loss can be written as:
 \begin{multline}
\label{eq:logloss}
-\mathbb{E}_{x,y \sim P_{\mathbf{X},\mathbf{Y}}}\big[\log v[x](y)\big] = \\
-\mathbb{E}_{x,y \sim P_{\mathbf{X},\mathbf{Y}}}\big[\log p(y\vert x)\big]
+ \mathbb{E}_{x,y \sim P_{\mathbf{X},\mathbf{Y}}}\!\left[\log\frac{p(y\vert x)}{v[x](y)}\right].
\end{multline}
The first term on the right hand side is the Shannon conditional entropy $H(\mathbf{Y}\vert \mathbf{X})$, which represents
the irreducible uncertainty about the task variable $\mathbf{Y}$ given the image data $\mathbf{X}$. The second term
is the expected Kullback--Leibler (KL) divergence between the true posterior and
the predictor, which is denoted as:
\begin{equation}
\mathbb{E}_{x,y \sim P_{\mathbf{X},\mathbf{Y}}}\!\left[\mathrm{KL}\!\left(p(y \vert x)\,\|\,v[x](y)\right)\right].
\end{equation}
This term quantifies how much additional uncertainty is introduced because the
observer’s predicted probabilities deviate from the true posterior.

Minimizing (\ref{eq:logloss}) over the predictors $v[x](y)$ in $\mathcal{V}$ and noting (\ref{eq:conditional-v-entropy})   yields:
\begin{equation}
\begin{aligned}
H_{\mathcal{V}}(\mathbf{Y}\mid \mathbf{X})
&= H(\mathbf{Y}\mid \mathbf{X}) \\
&\hspace{-2em}
+ \inf_{v\in\mathcal{V}}
\mathbb{E}_{x,y \sim P_{\mathbf{X},\mathbf{Y}}}
\Big[
\mathrm{KL}\!\left(
p(y\mid x)\,\|\,v[x](y)
\right)
\Big]
\end{aligned}
\end{equation}

This expression makes clear why $\mathcal{V}$-info can continue to change
even when rank-based measures such as AUC have saturated. Once an observer is
able to correctly order cases, further improvements in the quality of the
posterior, such as sharper probability estimates, better uncertainty
quantification, or reduced ambiguity, do not affect the ordering and therefore
do not change AUC. However, these improvements reduce the KL divergence term
above, leading to a decrease in $H_{\mathcal{V}}(T\mid Y)$ and a corresponding
increase in $\mathcal{V}$-info. Thus, while AUC reflects whether task
outcomes can be separated, $\mathcal{V}$-info reflects how completely and
reliably the image resolves uncertainty about the task for the observer class.
}

\end{document}